\documentclass[12pt,a4paper]{article} 
\usepackage[utf8]{inputenc}
\usepackage[english]{babel}
\usepackage{amsmath}
\usepackage{amsfonts}
\usepackage{amssymb}
\usepackage{graphicx}
\usepackage[font=footnotesize,labelfont=bf]{caption}
\usepackage{authblk}
\usepackage[colorlinks=true,citecolor=blue]{hyperref}
\usepackage{subfigure}

\usepackage[]{natbib} 
\usepackage{color}
\newcommand{\tr}{\color{red}}

\begin{document}
\renewcommand{\abstractname}{\vspace{-2\baselineskip}}

\title{Implosive collapse about magnetic null points:
\\ 
A quantitative comparison between 2D and 3D nulls}
\author[1,2]{J.~O.~Thurgood\textsuperscript{*}}
\author[1]{D.~I. Pontin}
\author[2]{J.~A.~McLaughlin}
\affil[1]{Division of Mathematics, University of Dundee, UK.}
\affil[2]{Department of Mathematics, Physics and Electrical Engineering, Northumbria University, UK.}
\affil[]{jonathan.thurgood@northumbria.ac.uk}
\date{}
\setcounter{Maxaffil}{0}
\renewcommand\Affilfont{\itshape\small}
\maketitle

\abstract{
    \textbf{
    	Null collapse is an implosive process whereby MHD waves focus their energy in the vicinity of a null point, forming a current sheet and initiating magnetic reconnection.
	We consider, for the first time, the case of collapsing 3D magnetic null points in nonlinear, resistive MHD using numerical simulation, exploring key physical aspects of the system as well as performing a detailed parameter study.
	We find that within a particular plane containing the 3D null, the plasma and current density enhancements resulting from the collapse are quantitatively and qualitatively as per the 2D case in both the linear and nonlinear collapse regimes.
	However, the scaling with resistivity of the 3D reconnection rate - which is a global quantity - is found to be less favourable when the magnetic null point is more rotationally symmetric, due to the action of increased magnetic back-pressure.
	Furthermore, we find that with increasing ambient plasma pressure the collapse can be throttled, as is the case for 2D nulls.
	We discuss this pressure-limiting in the context of fast reconnection in the solar atmosphere and suggest mechanisms by which it may be overcome.
	We also discuss the implications of the results in the context of null collapse as a trigger mechanism of Oscillatory Reconnection, a time-dependent reconnection mechanism, and also within the wider subject of wave-null point interactions.
	We conclude that, in general, increasingly rotationally-asymmetric nulls will be more favourable in terms of magnetic energy release via null collapse than their more symmetric counterparts.
    }
}
\vspace*{1cm}
\linebreak
{\tr{Manuscript in press, accepted for publication by The Astrophysical Journal as of February 2018. The final published version will be available with \lq{gold}\rq{} open access (available to all from Day 1), and should be considered the preferred version once available.}}
\newpage

 \section{Introduction} \label{sec:intro}
 
Magnetic reconnection is an important process for energy conversion throughout astrophysical plasmas, in the Sun and planetary magnetospheres as well as further afield -- for example in $\gamma$-ray bursts \cite[e.g.][]{zweibel2009,pontin2012}. {One particular location in which reconnection can preferentially take place is in the vicinity of a magnetic null point (locations at which the magnetic field strength is zero). Recent studies show that such null points exist in abundance in the solar atmosphere \citep{regnier2008,longcope2009}, and there is growing evidence that reconnection at nulls plays an important role in solar flares, CMEs, jets and bright points \citep[e.g.][]{barnes2007,2013ApJ...778..139S,morenoinsertis2013,2016ApJ...827...27Z,2017Natur.544..452W,2017A&A...605A..49C}. This is further supported by observations of flare ribbons that appear to result from particle acceleration during null point reconnection \citep{zuccarello2009,liu2011}.}

One particular mechanism involving reconnection at nulls that may be important for energy conversion is \lq{null point collapse}\rq{}: an implosive process whereby MHD waves
form large current densities by concentrating flux at small scales. If the ambient plasma pressure is sufficiently low, the collapse is thought to be halted at a scale determined by resistive diffusion, yielding  favourable scalings for  reconnection rates with decreasing plasma resistivity 
\citep{1992ApJ...393..385C,1996ApJ...466..487M}.
	The basic idea that  perturbations tend to collect at X-points (or X-type neutral lines) leading to a growth in current densities was first realised by \citet{dungey53}. 
	Null point collapse in 2D and 3D has subsequently  been considered from various perspectives. Dynamic collapse studies in the close vicinity of the null (within which the magnetic field and flow can be approximated as linear) have been performed in both the linear and nonlinear MHD regimes by \citet{Imshennik1967,bulanov1984,klapper1996,mellor2003}. 
	These studies tend to indicate unbounded growth of the current in the absence of dissipation. 		
	However, since they explicitly exclude the surrounding field, it is unclear whether sufficient energy could accumulate at the null in the full system to sustain this current blowup. 

	In this paper we take a different approach, which is to study the process computationally including the full nonlinear field and plasma flow geometries.
	This approach received significant attention during the 1990s in advocating null collapse as a possible mechanism for obtaining fast reconnection rates in two dimensions \citep{1991ApJ...371L..41C,1992ApJ...399..159H,1992ApJ...393..385C,1993ApJ...405..207C,1996ApJ...466..487M,2000mare.book.....P}, but is yet to be considered in 3D.
	In the 2D case, it was eventually realised that for the solar corona the plasma pressure would be likely sufficient to restrict the process and so questions were raised over its viability as a fast reconnection mechanism, at least from a solar physics perspective \citep[e.g.][]{1996ApJ...466..487M,2000mare.book.....P}. A possible resolution is hinted at in the work of \citet{1996ApJ...466..487M} in that nonlinear effects may overcome this limitation, permitting a secondary stage of fast reconnection after the initial halting, although to our knowledge this possibility has not since been investigated. More recently, PIC simulations of 2D null collapse find fast rates occur due to collisionless effects \citep{2007PhPl...14k2905T,2008PhPl...15j2902T}, and that null collapse seems to be able to efficiently accelerate particles which has been proposed as a source of $\gamma$-ray flares in the Crab Nebula \citep{2016arXiv160305731L}. However, since these simulations begin with a magnetic field in which the null is \lq{pre-collapsed}\rq{} at a kinetic scale, there remains the question of whether an external perturbation would initiate a collapse that forms a thin-enough current sheet to promote collisionless reconnection before being pressure-limited at an MHD-scale. 

An alternative perspective on null collapse is obtained by making  arguments based on lowest energy states in a domain in which the magnetic field lines are line-tied at the boundaries. Starting from an equilibrium, one perturbs the magnetic field in such a way as to displace some of the separatrix (2D) or spine and fan (3D) field lines at the boundaries, and then considers the properties of the new equilibrium that is accessible through an ideal dynamics with these boundaries held fixed. 
Studies using different analytical and computational approaches indicate that the lowest energy state contains a singular current layer at the null in both 2D and 3D \citep{syrovatskii1971,craiglitvinenko2005,pontincraig2005,fuentes2012,fuentes2013}. Thus, in any real plasma with finite magnetic Reynolds number, resistive effects must eventually become important during the collapse, making 2D and 3D nulls favourable sites for reconnection. Notably, pressure forces are shown to weaken the scaling of the singularity, but cannot remove it \citep{craiglitvinenko2005,pontincraig2005}.

In a recent study \citep{2017ApJ...844....2T}, we used MHD pulses to initialise a \textit{localised} and \textit{finite-duration} collapse of 3D null points, to form a current sheet embedded within a larger-scale (global) field and discovered that a phenomenon known as \textit{Oscillatory Reconnection} can occur at 3D nulls which are undergoing spine-fan reconnection \citep{pontinbhat2007a}. We noted that a number of qualitative  aspects of dynamic null collapse as described in the papers above appeared to carry-over to the case of 3D null collapse, but were at the time unable to quantitatively investigate the collapse phenomena  by way of parameter study due to restrictive computational requirements associated with those particular simulations.  
Thus, in this paper it is our primary aim to present the results of such a study and investigate the extension of known null collapse scalings 
{with resistivity},
 and associated magnetic reconnection
 {efficiency}, to 3D for the first time. 
	We also aim to present a fresh perspective on both 2D and 3D null collapse which suggests that it may in-fact be a phenomenon of some importance even in the case of relatively \lq{high}\rq{} coronal back-pressures.  
The paper is structured as follows; first, we out outline the set-up of our numerical experiments (section \ref{setup}). Then, in Section \ref{review} we briefly discuss  physical aspects of null collapse based on past 2D results and the qualitative similarity  and differences in extension to 3D. 
	We then in Section \ref{scaling} present 
	{the scaling of key measurements with variable resistivity} in 2D and 3D collapses in low-$\beta$ plasmas as, calculated from our simulations, in order to investigate empirically the applicability of previous analysis to the 3D case. 
We then consider the possibility of null collapse as a fast reconnection mechanism in astrophysics, and the effects of more appreciable ambient plasma pressure in Section \ref{sec:higherbeta}. 
We summarise our results and draw conclusions in (Section \ref{discussion}).

\section{Simulation Setup}\label{setup}

The {simulations} involve the  numerical solution {of the 2.5D and 3D} single-fluid, resistive MHD equations using the LareXd code \citep{2001JCoPh.171..151A}. Here we outline the simulation setup (initial conditions), with full technical details deferred to the appendix. All variables in this paper are nondimensionalised
{(see appendix \ref{appendixA})},
unless units are explicitly stated.

\begin{figure*}
	    \centering
    \includegraphics[width=0.875\linewidth]{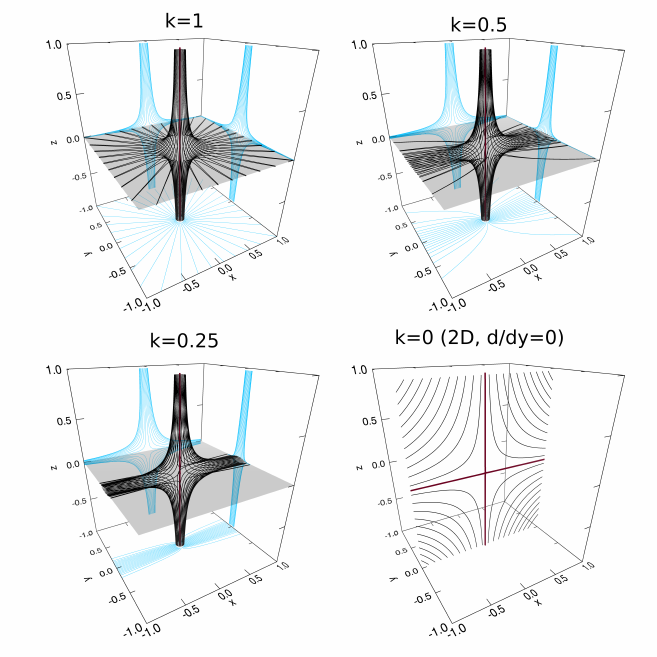} 
    \caption{Representative fieldlines about the 3D equilibrium magnetic fields $\mathbf{B}_{0}$, for variable eccentricity parameter $k$  (Equation \ref{eq:3dnull}) . In the $k=1$ case (a \lq{proper}\rq{} null point) topologically consists of a spine fieldline (red line), running along the $z$-axis into the null (at the origin), and the fan plane (transparent grey), sitting in the $z=0$-plane, in which fieldlines run radially outward from the null. Elsewhere, fieldlines run hyperbolically, parallel to the spine or fan when far from the null. As $k$ is decreased, the fieldlines become increasingly eccentric and take on a prefered direction in the fan plane. At $k=0$, the null reduces to a translationally invariant 2D null structure, which is a $45^{{\circ}}$ rotation of Equation \ref{eq:2dnull}, consisting of two sets of separatricies (red fieldlines). In the $k\neq0$ cases, the light blue fieldlines illustrate planar projections of the selected fieldlines. 
    }
 \label{fig:fieldlines}
\end{figure*}

We consider the collapse of {magnetic fields containing null points,}  of the 2D Cartesian form 
\begin{equation}\label{eq:2dnull}
\mathbf{B}_{0}=\left[z,0,x\right],
\end{equation}
and the 3D form 
\begin{equation}\label{eq:3dnull}
\mathbf{B}_{0}=\left[x,ky,-(k+1)z\right],
\end{equation}
{each of which is a potential field, free from electric currents, and so constitutes a minimum energy state.} 
	{These fields are therefore force free ($\mathbf{j}\times\mathbf{B}_{0}=\mathbf{0}$)}.
{These prescriptions, often referred to as \lq{linear null points}\rq{}, arise from the first-order Taylor series expansion near a null point embedded within a generic field,  and so represent an approximation to  realistic fields sufficiently close to the null point itself \citep[see, e.g.,][for full details]{1996PhPl....3..759P,2000mare.book.....P}. }
	We present results for four different geometries in this paper, namely the 2D null (equation \ref{eq:2dnull}) in conjunction with enforcing ${\partial}/{\partial}y=0$ throughout the solution (reducing the equations to so-called 2.5D) and the 3D null (equation \ref{eq:3dnull}) with the fully 3D MHD equations for $k=0.25$, $k=0.5$, and $k=1$. 
		Note, that setting $k=0$ in equation (2) and rotating (the transformation $x{\rightarrow}z$, $z{\rightarrow}-x$) recovers the same 2D null point as equation (\ref{eq:2dnull}).
		In practice,  we use the form of equation (1) preferentially in the 2D simulations for computational reasons,  although the underlying physical problem is essentially the same.
	{The fieldline structure of these fields is illustrated in Figure \ref{fig:fieldlines}. Topologically, the fields consist of the  null point itself  (located at the origin), and (in 2D) a set of separatrices, or (in 3D) a spine field line, running along the z-axis toward the null point, and a fan plane z = 0, consisting of field lines pointing radially outward. Other field lines, connectively separated by the separatrices (or spine and fan in 3D), have a hyperbolic structure. The parameter $k$ is an eccentricity parameter, introducing an azimuthal asymmetry to these fieldlines about the $z$-axis, taking on a preferential  direction within the fan plane (the $x$-direction), and an associated rescaling of magnetic field strength in the $z$-direction}.

 {The computational domain is the cube $|x,y,z|\le1$ (or equivalent square in 2D), } with the boundaries  taken to be closed with a fixed { magnetic} through-flux. The ratio of specific heats is $\gamma=5/3$ throughout.  
 	Plasma resistivity $\eta$  is taken as uniform and is a variable of the parameter study,
 	{and under our normalisation its value corresponds to an inverse Lundquist number as defined by the normalisation constants (see Appendix \ref{appendixA} for further details).} 
	{Given that our outer boundaries are located at a nondimensional distance of 1 from the null, where the local Alfv\'en speed is of the order unity under the normalisation, this Lundquist quantifies the relative strength of the diffusivity of the whole domain.
	}
	 We take the plasma to be initially at rest ($\mathbf{v=0}$), of uniform density ($\rho=1$) and a uniform gas pressure, chosen such that a fixed plasma-$\beta$ defined at radius $r=1$ may be set as a variable, denoted $\beta_0$. 
	{{Thus, the background state of the simulations is given by the force-free fields (equations \ref{eq:2dnull}, 2D, and \ref{eq:3dnull}, 3D) and a uniform gas pressure (zero pressure gradient), and is therefore  an exact equilibrium (it has furthermore been verified to be a numerical equilibrium, as described in the Appendix \ref{boundary}).  }}

	{{This background state is then}} subject to finite amplitude perturbations  to {the total field} $\mathbf{B}=\mathbf{B}_{0}+\mathbf{B}^{\prime}$  of the Cartesian form
\begin{equation}\label{eq:2dperturbation}
\mathbf{B}^{\prime} =\frac{j_{0}}{2}\left[z,0, -x\right]
\end{equation}
 and in 3D,
\begin{equation}\label{eq:3dperturbation}
\mathbf{B}^{\prime} =j_{0} \left[0,0,x \right].
\end{equation}
	The perturbation, given by equations \ref{eq:2dperturbation} and \ref{eq:3dperturbation}, corresponds to a superimposed, uniformly distributed current density in the $y$-component $j_{y}$ of initial magnitude $j_{0}$. 
 This current means that the separatrices (or the equivalent spine fieldline and fan separatrix plane in 3D) are no longer perpendicular.
	This {{ perturbation to the initial background}} therefore {{ disrupts}} force balance and, immediately after initialisation, the {{ perturbation begins to 
focus towards the null, establishing a plasma flow that drives the collapse, as described qualitatively in the following section.}}
	{{We note that, due to the nature of our boundary conditions (Appendix \ref{boundary}), that this corresponds to a  perturbation which although uniform in the domain extends to the boundaries only. The boundary conditions are ${\bf v}={\bf 0}$ and ${\bf B}$ line-tied (normal component fixed). As such there is no inflow of either plasma or magnetic energy. This is a crucial difference between our study and the linear collapse studies of e.g.~\citet{Imshennik1967,bulanov1984,klapper1996}: in our case there is no energy inflow through the boundaries to drive unlimited collapse. This perturbation is thus of fixed spatial extent  (extending to the edge of the domain) and finite total energy.  The effects of boundary conditions on collapse, which is a subtle issue, is discussed further in Chapter 7.1 of \citet{2000mare.book.....P} and also by, e.g.,  \citet{1979JPlPh..21..107F,1998PhPl....5..910K}. As such, we stress the overall setup in this paper is that of a \textit{finite perturbation} to a \textit{stable system}. }}

{{ Presupposing this localised} initial disturbance {{to the magnetic flux}} is motivated by the well-established result that MHD waves are generically attracted to null points, and so we expect that external perturbations to the larger-scale magnetic field will preferentially collect at nulls \citep[see][for a review]{2011SSRv..158..205M}. Specific examples of this in application can be seen in the work by \citet{2015A&A...577A..70S} and \citet{2017ApJ...837...94T} where photospheric motions lead  to current accumulation at nulls in realistic model solar atmospheres. In our paper, where we focus on the details of the current sheet formation (i.e. dynamics very close to the null rather than those in the external field), we thus presuppose this disturbance as  both a matter of computational feasibility  and as a modelling simplification (in line with previous null collapse studies, e.g. \citealt[][]{1996ApJ...466..487M,2007PhPl...14k2905T}).}

\section{A brief overview of the physical and qualitative aspects  of 2D and 3D null collapse in low-$\beta$ plasma}\label{review}

We begin by outlining the qualitative and physical aspects of our various simulations,  before going on to quantify these results in Section \ref{scaling}. The perturbations considered in this paper (Section \ref{setup})  correspond to an initial enhancement of free magnetic energy centred around the null. 
	The magnetic field is not in force balance, and after $t=0$ the evolution can be understood in terms of the propagation of the perturbation towards the null  as a MHD wave. 
	Due to the arrangement of the Lorentz force, the incoming wave immediately drives a fluid flow typical of reconnection, with the null itself coinciding with a stagnation point separating symmetric and anti-symmetric  regions of inflow and outflow, delineated by the separatrices, or the spine and fan in 3D  (i.e. streamlines of this flow field resemble rectangular hyperbolae). 
	It is the focusing of this incoming wave (namely, its excess magnetic flux, and the flow it forces through its Lorentz force, both of which will increase in magnitude in time) which is at the heart of null collapse.

	In the $\beta<1$ case here, the incoming perturbation propagates  as  a fast (magnetic) MHD wave. 
	Since first discussed by \citet{dungey53}, it has become  well-established that such waves are, in-general, attracted to null points due to a refraction effect close to nulls, with fast waves  propagating across surfaces of constant Alfv\'en speed, or down the potential well of the background field  \citep[e.g.][for a review]{2011SSRv..158..205M}. 		
	Accordingly, the energy  of the perturbation is propagated in towards the null point, transporting its flux/magnetic energy and associated mass-flow. 
	As such, null collapse is a class of MHD implosive process with a null point being the center of converging \textit{magnetic flux},  and also the aforementioned  converging-diverging flow (in this sense, it differs somewhat to most implosions which involve only a convergence of mass at a symmetry point, line or plane, dependent on dimensionality).
	In linear MHD such an implosion evolves in a relatively simple manner. Close to a generic null the field grows linearly away from the null: external disturbances accumulate at the null  according to the linear Alfv\'en speed profile, and therefore the volume in which the energy is contained decreases exponentially (viz. gradients across this volume will increase exponentially).  {Therefore, as perturbations focus their energy in the vicinity of the null }an exponential increase in current density at the null point occurs, a result that has been demonstrated both in the null collapse literature \citep[e.g.][]{1992ApJ...393..385C,1996ApJ...466..487M} and in papers which consider current build up due to more generalised, externally originating MHD waves \citep[e.g.][]{2004A&A...420.1129M,2007JGRA..112.3103P,2011SSRv..158..205M,2012A&A...545A...9T}. 
	This indicates that, at least in the low-$\beta$ limit, the process is not too dependent on the rather symmetric initial conditions often employed in the collapse studies. 
	As the waves cannot reach the null by propagation at the background Alfv\'en speed  ($v_{A}\rightarrow0$ as $r\rightarrow0$)  this focusing is expected to continue until resistive diffusion and/or plasma back-pressure becomes sufficiently large to allow the perturbations to propagate/diffuse through to the null, possibly reflecting the wave.
	 { In the related case of the collapse of a 2.5D X-line, the process is essentially the same although the guide-field also provides a magnetic back-pressure which may oppose the collapse, in a manner analogous to the plasma back-pressure \citep[][]{1996ApJ...466..487M}.  }

	In the more physically realistic case of a finite-amplitude (nonlinear) perturbation, the increasing amplitude of the focussing wave means that the excess flux carried by the wave may
overwhelm the background field, so that the wave undergoes nonlinear {evolution}.
	Indeed, this will inevitably occur unless the collapse is first halted by resistive diffusion and back-pressure as described above.
	As such, for our nonlinear MHD simulations, we expect there to be two distinct regimes in which the implosion evolves, depending on the initial wave energy. If the perturbation energy is sufficiently large that nonlinear {evolution} occurs, the implosion becomes characterised by quasi-1D behaviour (and as we will see in Section \ref{scaling}, different scaling laws as a consequence). 
	This is because the magnetic field of the perturbation reinforces the background field in two quadrants -- where the wavefront undergoes nonlinear {evolution} -- and partially cancels it in the other two quadrants, where the wavefront stalls \citep{1992ApJ...393..385C,1996ApJ...466..487M,2011A&A...531A..63G}.  
	This self-reinforcing process means that  nonlinear collapse naturally creates sheet-like current distributions, as we will soon demonstrate. 
	The quasi-1D phase of the collapse is closely related to the case of imploding 1D Harris-like current sheets (i.e., anti-parallel field about a null line) such as those considered by \citet{1982JPlPh..27..491F} and \citet{2015ApJ...807..159T}, where even, in certain limits, previous 2D null collapse and 1D collapse solutions have been shown to be equivalent \citep[see e.g., the Appendix of][]{1982JPlPh..27..491F}.

\begin{figure*}
	    \centering
    \includegraphics[width=0.875\linewidth]{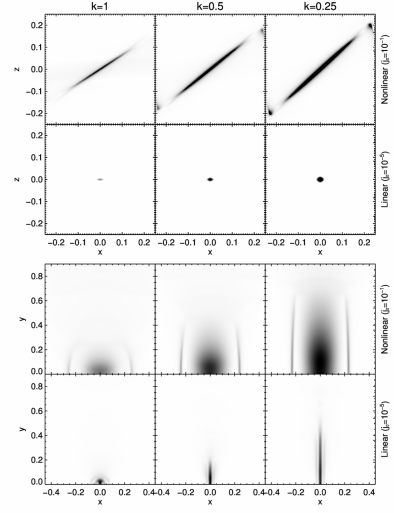} 
    \caption{Current sheet morphology at peak time for 3D, low-$\beta$ collapse in the linear and nonlinear regimes, where here $\eta=3\cdot10^{-4}$. Top:
      $\int_{-1}^{1} |j(x,y,z)| \,dy$ illustrating the current distribution in the $x-z$ halfplanes (a \lq{}side-on\rq{} view of the null), where the top row corresponds the nonlinear regime ($j_{0}=10^{-1}$) and the bottom the linear ($j_{0}=10^{-1}$). Colours are fixed to a constant, saturated, linear scale for each row.  We find that, in planes of fixed-$y$ sufficiently close to the null that the collapse is qualitatively as per the 2D case. Bottom: Corresponding distributions of $\int_{-1}^{1} |j(x,y,z)| \,dz$ illustrating the current distribution in the $x-y$ halfplanes (a \lq{}top-down\rq{} view of the null). We find that the enhanced magnetic pressure for increasingly rotationally symmetric nulls ($k\rightarrow1$) inhibits collapse away from the null by providing a magnetic back-pressure via the growing out of plane field component $B_{y}=ky$. 
    }
 \label{fig:qualitative3D}
\end{figure*}

We noted in our discussions of \citet{2017ApJ...844....2T}, where null collapse was used to trigger \textit{Oscillatory Reconnection},  that collapse for a $k=1$ 3D null appeared to proceed in a qualitatively similar way to the 2D collapse, within a particular plane containing the null point 
due to the same nonlinear {evolution} process of 2D collapse. We also showed (for a specific case) that the resulting reconnection was of the spine-fan type \citep[][Figure 4]{2017ApJ...844....2T}, which is consistent with related results of 3D reconnection excited by disturbances emanating at the computational boundary \citep[e.g.][]{pontinbhat2007a}. In the following section we aim to more broadly quantify this 3D collapse and the associated reconnection, for a variety of plasma parameters and null geometries. 
	We note that during the process, the spine and fan of the null point locally collapse towards one another (see Figure \ref{fig:qualitative3D}). The plane in which this collapse occurs (the plane that contains the deformed spine line) is determined both by the perturbation that drives the collapse and by the null-point structure. In the remainder of this paper, we refer to this plane as the \emph{plane of collapse}.

	Figure \ref{fig:qualitative3D} shows some example current sheets resulting from this parameter study. 
	It shows integrated values of the current density through  the $xz$-planes (\lq{side-view}\rq{}) which illustrates  the aforementioned similarities to 2D collapse in planes of fixed-$y$, and also integrated values through $xy$-planes (\lq{top-view}\rq{}) which shows the out-of-plane distribution of current. 		
	Importantly, as $k$ is increased from $k=0$ (the 2D null with no guide-field) to $k=1$ (the rotationally symmetric limit), the out-of-plane extent of the current cylinder/ring (linear regime) or sheet (nonlinear) is reduced by a factor commensurate with the change in $k$ -- consistent with the behaviour found in boundary-driven simulations by \citet{alhachami2010,galsgaard2011b}. This is due to the influence of the magnetic field component, $B_{z}=ky$, which provides a magnetic back-pressure to resist the collapse (away from $y=0$) in a manner analogous to the guide-field in the 2.5D X-line case \citep[e.g.][]{1996ApJ...466..487M}.  
	We will see that this has an important impact on the overall reconnection rate. 
	These current concentrations can also be further contextualised in terms of the collapse field by comparison of 	Figure \ref{fig:qualitative3D}  to \citet[][Figure 2b]{2017ApJ...844....2T}. 

\section{Quantitative scaling in  low-$\beta$, compressible plasma}\label{scaling}

We now present the results of a parameter study of collapsing null points of the form given by equations (\ref{eq:2dnull}) and (\ref{eq:3dnull}) in a low pressure plasma such that $\beta_{0}=10^{-8}$, for a variety of uniform plasma resistivity values in the range $10^{-4}{\,}{\leq}{\,}{\eta}{\,}{\,}{\leq}10^{-2}$ 
{(i.e. global Lundquist numbers in the range $10^{2}-10^{4}$)}, and (in 3D) for field line eccentricities $k=0.25, 0.5, 1.0$. 
The data shown in this section is for such simulations where the nulls are subjected perturbations (equations \ref{eq:2dperturbation} and \ref{eq:3dperturbation}) which correspond to (initially) uniform current densities of magnitude $j_{0}$ in the  $j_{y}$ component.  We consider two different fixed-energy perturbation amplitudes, the first ($j_{0}=10^{-5}$) is expected to reside in the linear regime for the range of $\eta$ considered, and the second ($j_{0}=10^{-1}$) is expected to behave nonlinearly, according to the arguments of \citet{1993ApJ...405..207C}.
	In the following subsections we first consider the critical times (the halting times) of each implosion, then present measurements of current sheet geometries along the width-wise, length-wise, and out-of plane (3D) axes, then finally the  peak reconnection rates and evaluate also the efficiency of overall flux transfer by the critical time. 	
{Due to the extremely low resistivities of astrophysical plasmas, which are so small that they cannot be directly simulated, we pay particular attention to the scaling of these quantities across the computationally accessible range of resistivity. This mirrors the standard approach of previous (2D) collapse studies, such as \citet{1996ApJ...466..487M}. These scalings, by extrapolation, allow for an estimate of collapse behaviour as resistivity is further reduced.
}	%

\subsection{Critical times (time of peak reconnection)}

\begin{figure*}
    \centering
    \includegraphics[width=0.5\linewidth]{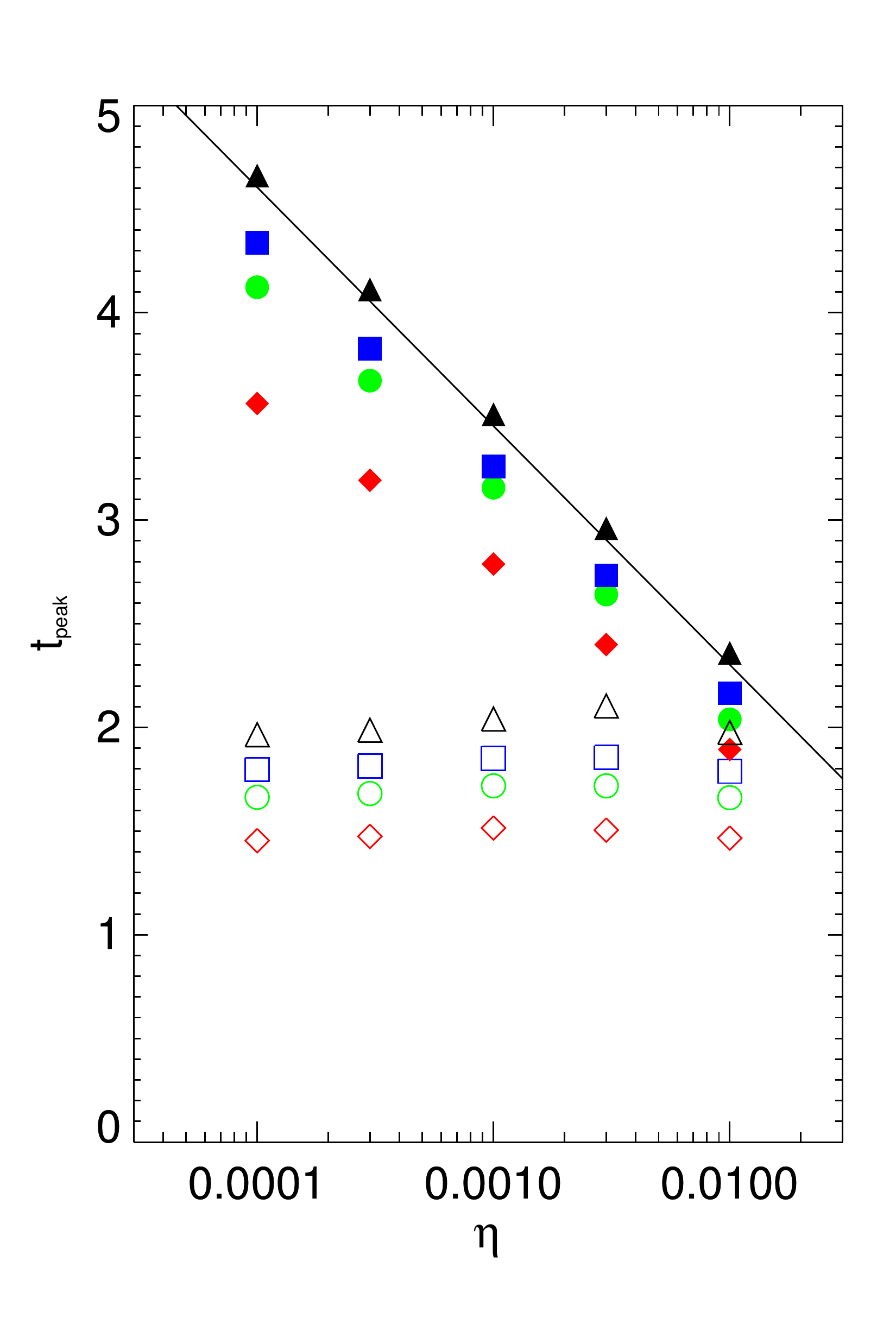}  
    \caption{Time of peak current density (i.e. {critical time}), for linear and nonlinear perturbations at 2D (triangles) and 3D nulls (for eccentricities $k=0.25,0.5,1.0$ with corresponding symbols; blue squares, green circles and red diamonds respectively). The solid line shows $t_{peak}=0.5 \ln|\eta|$, the theoretical (linear) wave travel time for a pulse at the boundary to reach the critical diffusion radius in the case of a 2D (or $k=0.0$ 3D) null point. The  filled shapes indicate runs with linear perturbation amplitude $j_{0}=10^{-5}$, and  the unfilled shapes are for nonlinear perturbation amplitude  $j_{0}=10^{-1}$. 
         }
         \label{fig:tpeak}
\end{figure*}

	Figure \ref{fig:tpeak} shows the time of peak current density (i.e. critical time, or the time of halting) as a function of plasma resistivity for different 2D and 3D runs in the linear and nonlinear collapse regimes. As pointed out by \citet{1992ApJ...393..385C}, a simple calculation suggests for the 2D null that the peak time in the linear case will simply be the time for the components of the perturbation at the boundary to travel to the critical radius $r_c$ of the diffusion region, which is $r_{c}=a\sqrt{\eta}$ where $a\sim1$. Thus, for the 2D null, the shortest such time is for a radially propagating mode, giving $t_{peak}=0.5 \ln|\eta|$, which is indicated by the straight line in the figure. The 2D simulations in the linear regime are in good agreement with this prediction, and we find that the principal effect of considering 3D nulls of decreasing eccentricity from the translationally invariant $k=0$ case to the rotationally symmetric $k=1$ case is a decrease in the peak time. This is naturally explained by the increasing background Alfv\'en speed as $k\rightarrow1$ (where $B_{0}^{2} = x^{2} + k^{2}y^{2} + \left[k+1\right]^{2}z^{2}$) decreasing the signal travel times. 
These critical times are found to be independent of the perturbation energy, so long as it is sufficiently small that the perturbation remains always in the linear regime (i.e., they are unchanged for other small values of $j_{0}$, such as $j_{0}=0.1{\eta}$, which we do not show here). This amplitude-independent peak time confirms the linearity of the  dynamics involved in these particular collapses. 
The perturbations which are expected to behave nonlinearly are indicated by unfilled symbols in Figure \ref{fig:tpeak}. 
Our simulations confirm that this is also the case for the 3D null collapses, and again, we find that the primary effect of decreasing 3D null point eccentricity is the uniform reduction in overall peak time across the range of $\eta$ considered. This is, like the linear case, simply a consequence of faster background Alfv\'en speeds at the 3D nulls (which still influences the collapse and the focusing of the pulse before it enters its nonlinear stage). 
We note that we also considered a set of runs with perturbation $j_{0}=10^{-2}$, which was be expected to transition between the linear and nonlinear collapses for the range of $\eta$ considered. We indeed found that such runs begin to depart from the straight line when $j_{0}$ becomes of the order $\eta$ for both 2D and 3D collapses in a manner reminiscent of \citet[][their Figure 1]{1996ApJ...466..487M}, which we have not shown here to avoid excess clutter in the figure.

\begin{figure*}
    \centering
    \subfigure[]{\includegraphics[width=0.4\linewidth]{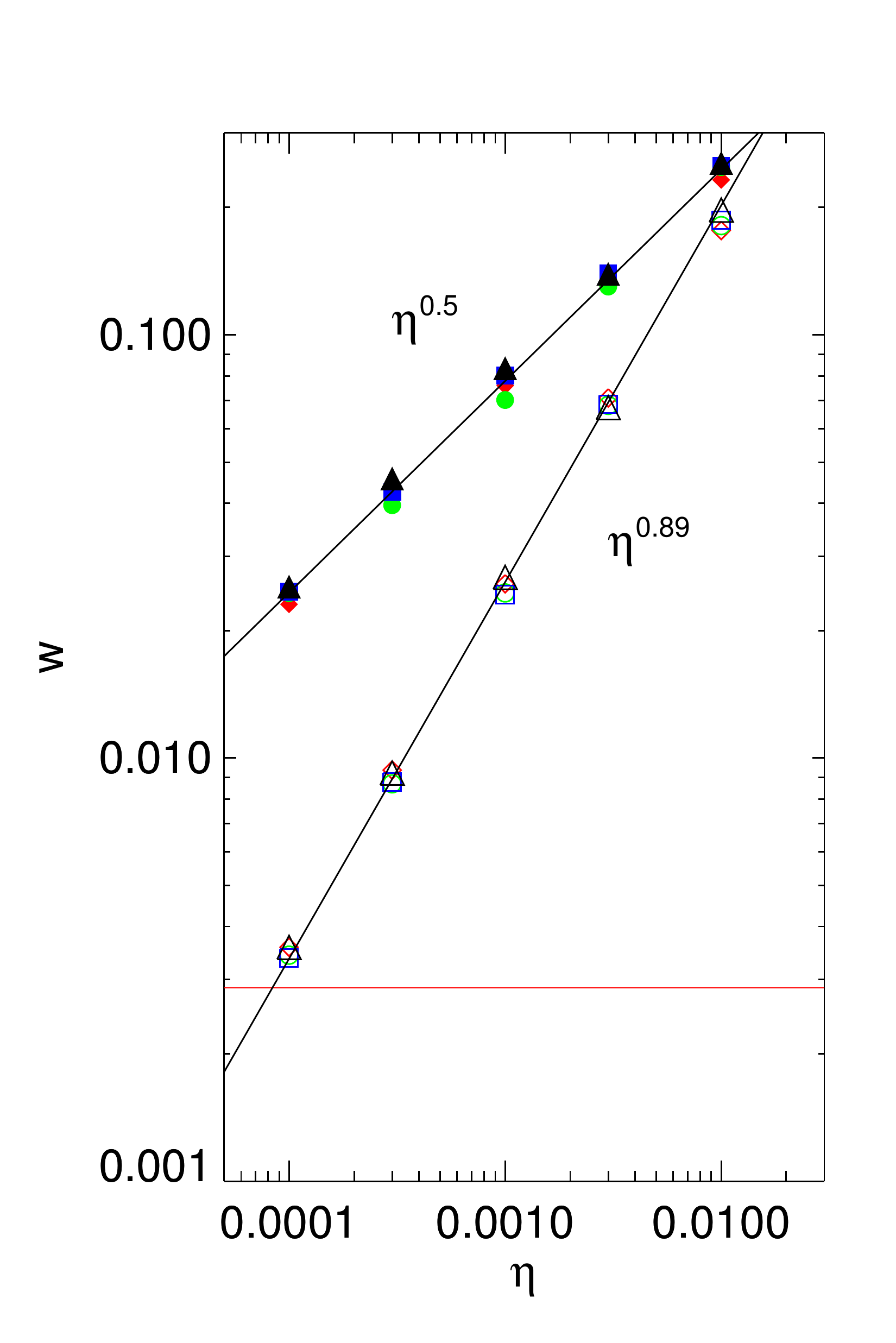} }
        \subfigure[]{\includegraphics[width=0.4\linewidth]{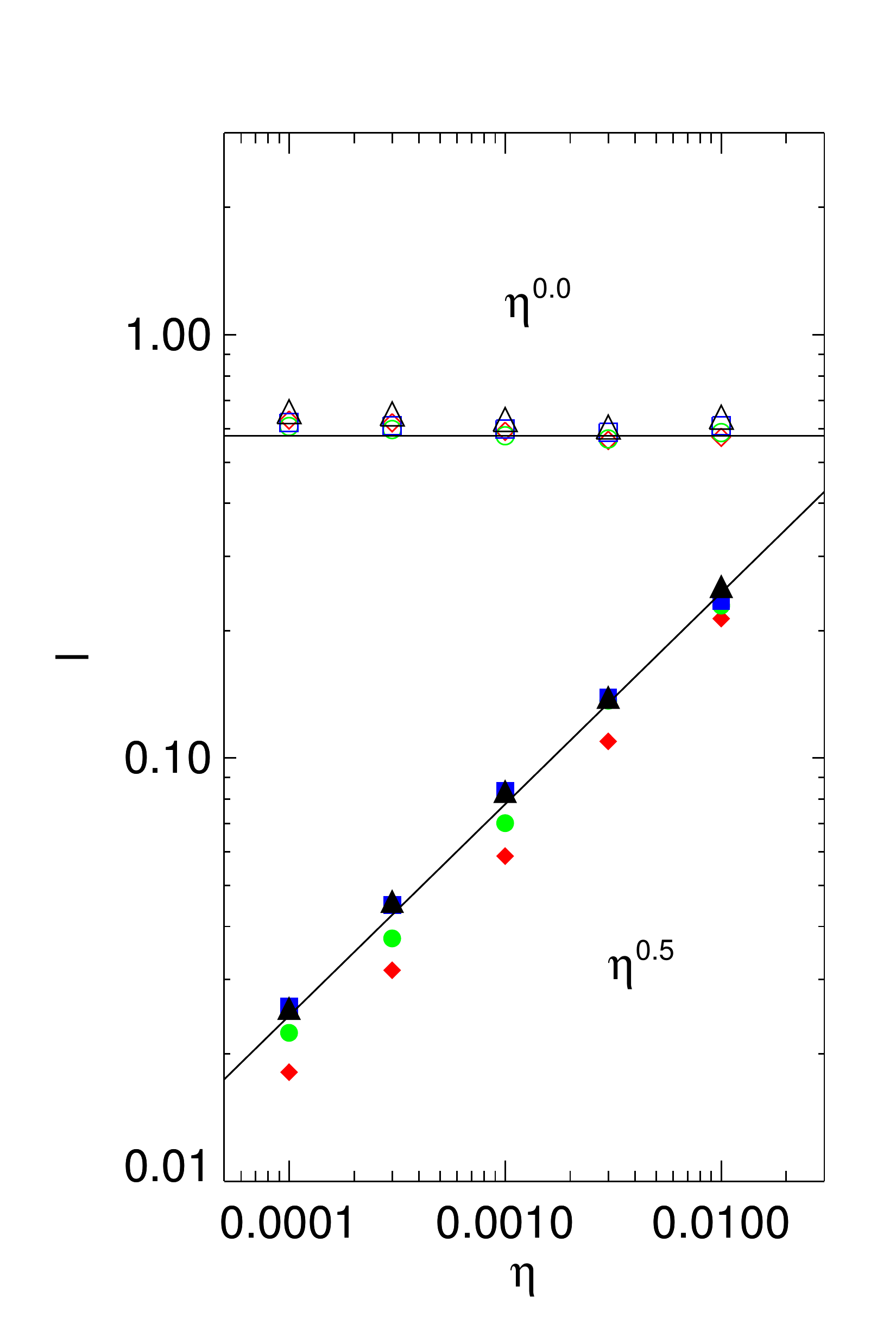} }

    \caption{
	Measured current sheet width ($w$, left) and length
     ($l$, right) at time of peak current, for the 2D (upward black triangles) and 3D collapses (for eccentricities $k=0.25,0.5,1.0$ with corresponding symbols; blue squares, green circles and red diamonds respectively) where the linear amplitudes correspond to the filled shapes and the nonlinear to the unfilled. 
     Black lines show typical scalings for comparison with the data.
     The red horizontal line indicates the size of $10{\Delta}x_{min}$ for the 3D simulations, indicating the resolution across the current sheets at this peak time. 
         }
         \label{fig:wlpeak}
\end{figure*}

\subsection{Current sheet geometry}

\begin{figure*}
    \centering
    \subfigure[]{\includegraphics[width=0.4\linewidth]{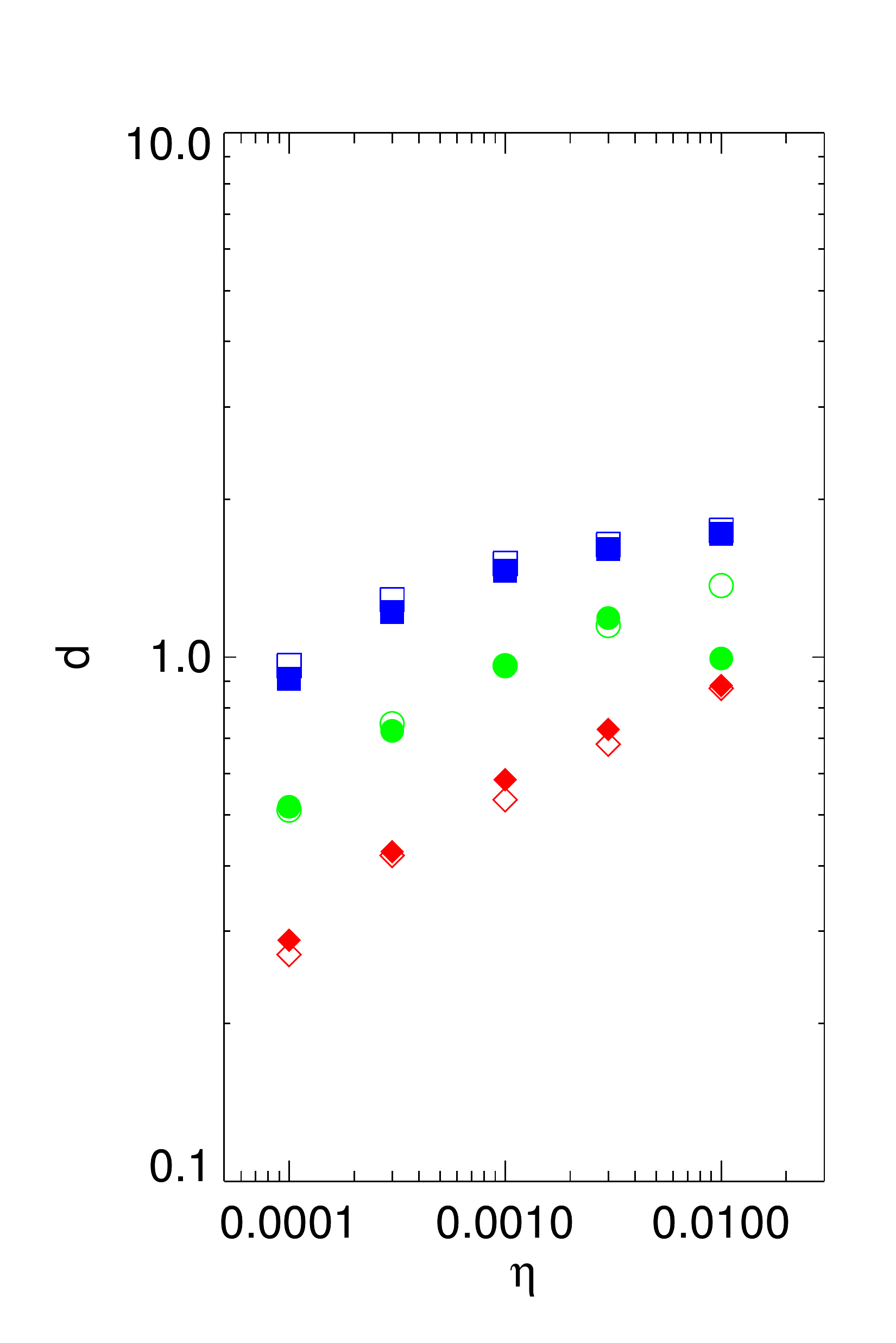} }
        \subfigure[]{\includegraphics[width=0.4\linewidth]{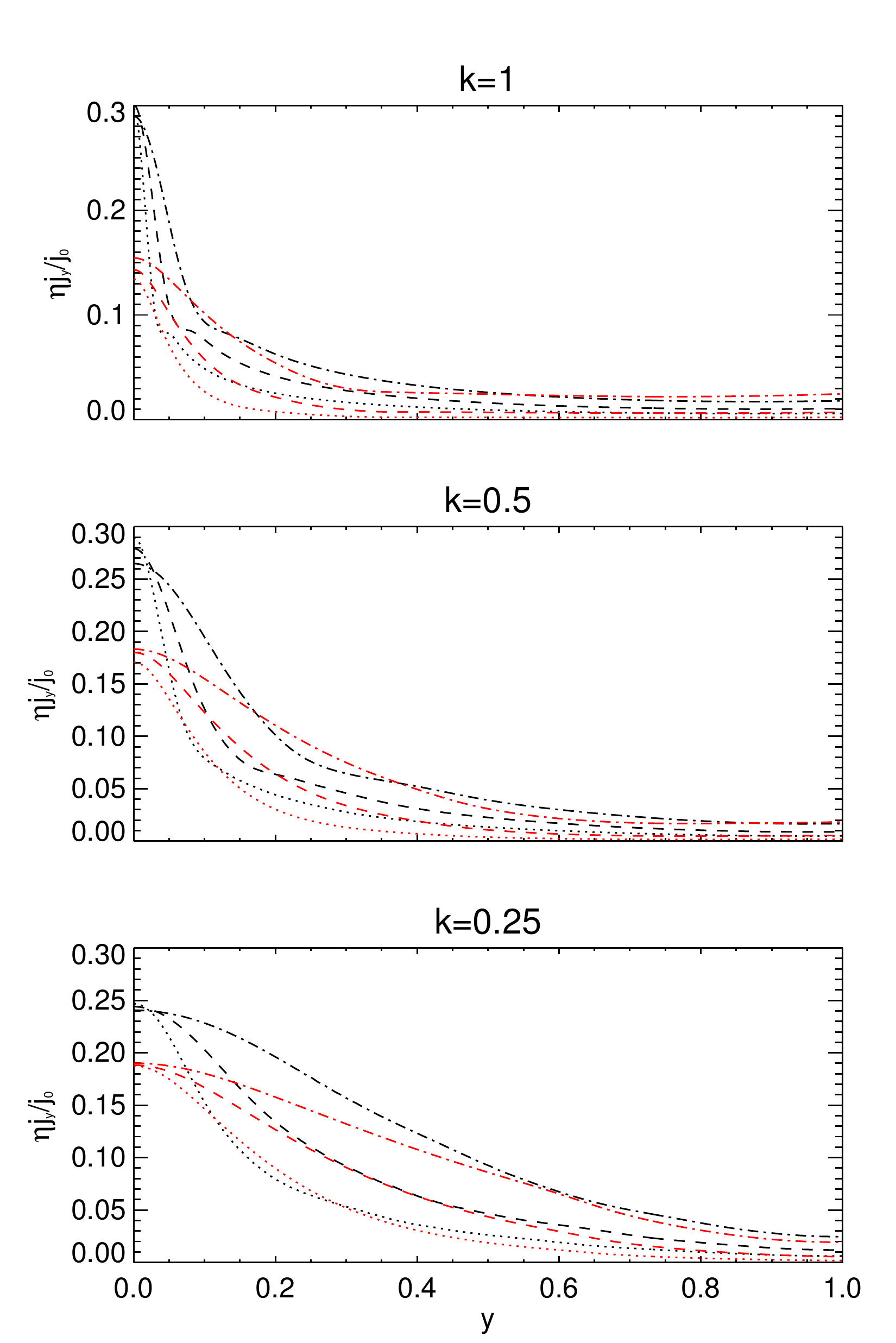} }

    \caption{Left: Measured current sheet \lq{depth}\rq{} $d$, for eccentricities $k=0.25,0.5,1.0$ with corresponding symbols; blue squares, green circles and red diamonds respectively. Between different values of $k$, the measured $d$ scales as approximately $k^{-1}$. Right: representative profiles a the peak time along the $y$-axis of the product ${\eta}j_{y}$, normalised to $j_0$,    for  resistivity ${\eta}=10^{-4}$, $3\cdot10^{-4}$, and $10^{-3}$ for dashed-dot, dashed and dotted lines, respectively, with black (red) lines corresponding to the linear (nonlinear) case.
         }
         \label{fig:dfit}
\end{figure*}

Next, we consider geometrical scalings of the current sheet formed at the critical/peak times identified. Figure \ref{fig:wlpeak} shows measurements of the current sheet width $w$ and length $l$ for the different linear and nonlinear runs. Here, with regards to 3D nulls, width and length are defined in the $y=0$ plane (the plane of collapse) by analogy to the 2D null, and we will consider the out-of-plane distribution shortly (current sheet \lq{depth}\rq{} $d$). As per Section \ref{review} for runs in the linear regime the current distribution is not a \lq{}true current sheet\rq{} but rather forms a more uniform ring-like distribution in the plane of collapse. For the 2D linear runs, we find that our simulations conform to the expected $\eta^{0.5}$ radial scaling discussed by \citet{1996ApJ...466..487M} with $w=l$. For the 3D linear runs, $w$ was (arbitrarily) chosen to lie along the $x$-axis and $l$ along the $z$-axis. The measurement conforms closely to the $\eta^{0.5}$ line with no discernible difference with eccentricity in the case of measurements along the x-axis ($w$), whereas the length $l$ along the $z$-axis uniformly decreased for decreasing eccentricity due to the $B_{z}{\neq}B_{x}$ imbalance. 
	Thus, linear collapse at decreasingly eccentric 3D nulls produces increasingly ellipsoidal current distributions within the plane of collapse , departing from the cylindrical symmetry of 2D due to the nonuniformity of the background Alfv\'en speed (this feature is visible in the second row of Figure \ref{fig:qualitative3D}).
	For the nonlinear runs, $w$ and $l$ are defined more meaningfully as the short and long current sheet axis (in the $y=0$ plane, for the 3D case). We find that for both our 2D and 3D runs, the width scales as $\eta^{0.89}$ (unfilled symbols, Figure \ref{fig:wlpeak}a). \citet[][their section 2.2 and 2.3]{1996ApJ...466..487M} proposed that 
this scaling should be $\eta^1$ in the absence of density inhomogeneities in the current layer, or $\eta^{0.89}$ if such inhomogeneities play a role in halting the collapse.
Their argument was based on comparison with 1D analytical results by \citet{1982JPlPh..27..491F} describing the ideal implosion of a planar current sheet (taking into account the formation of density inhomogeneities) to the scale at which the diffusion speed should become comparable to wavespeeds in the inflow region.  As such, this indicates that even in these initially low-$\beta$ simulations growing current sheet inhomogeneity plays a role in halting the collapse and setting the smallest width $w$ it can access -- although we stress that it is ultimately halted by diffusive effects, as opposed to, say, a back-pressure associated with adiabatic heating in the compressed current sheet which would be $\eta$-independent. Figure \ref{fig:wlpeak} also indicates that null point eccentricity has little influence on current sheet width. 
We also find that for the nonlinear runs, the measured current sheet length $l$ shows only a very weak (if any) dependence on $\eta$. This is consistent with the notion of \citet{1992ApJ...393..385C} that the length is set by the point at which nonlinear {evolution} occurs in the width-wise direction (and so controls length-wise stalling). This is ultimately determined by the energy of the imploding perturbation, which is fixed across the different nonlinear runs. There is some slight variation with eccentricity, but unlike for other parameters there is no clear pattern. As such, we do not attribute any specific physics to this variability.

\begin{figure*}
    \centering
    \subfigure[]{\includegraphics[width=0.4\linewidth]{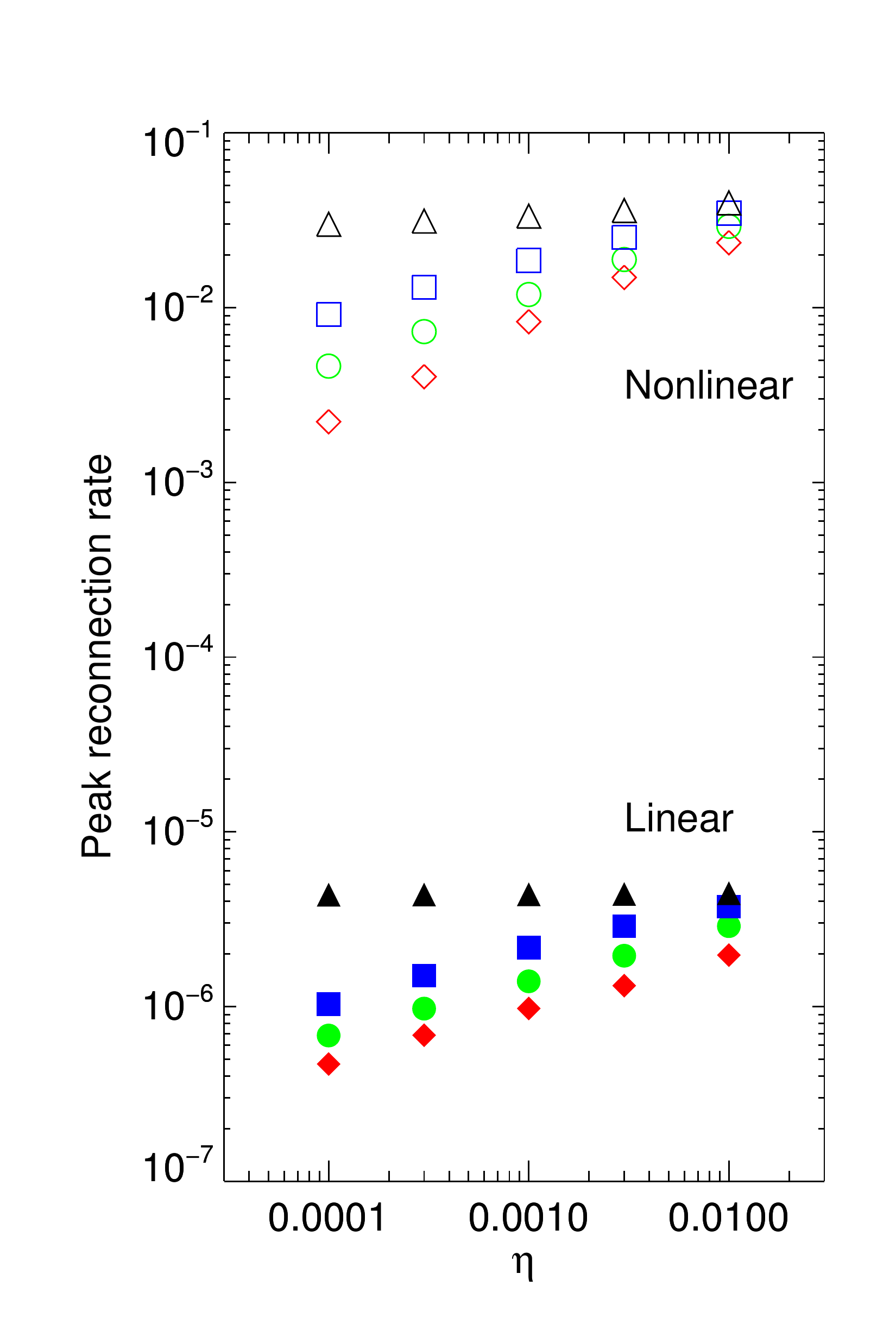} }
        \subfigure[]{\includegraphics[width=0.4\linewidth]{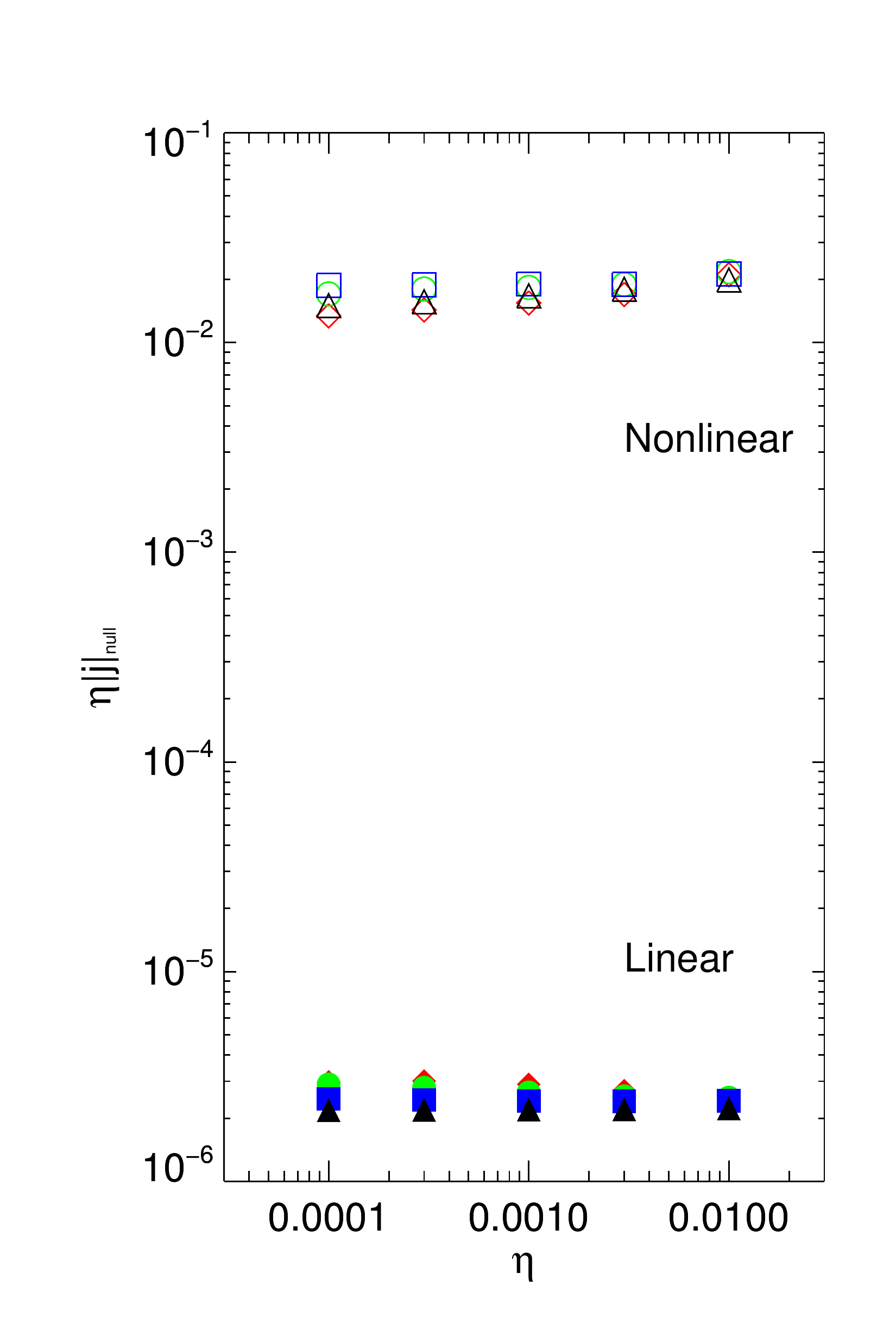} }
       \caption{Left: Measures of the peak 3D reconnection rate, for the 3D runs with $k=0.25$,$0.5$ and $1.0$ (blue square, green circle and red diamond, respectively). A comparative 3D reconnection rate for the 2D runs (black triangles) is obtained by multiplying the 2D reconnection rate by the equivalent out-of-plane length $L_y=2$. Right: The product ${\eta}j$ at the null itself, which is similarly enhanced by collapse in the plane of collapse regardless of $k$. 
         }      
 \label{fig:rates}
\end{figure*}

Turning now to the current sheet `depth' $d$ along the $y$-axis in the 3D simulations, the geometry of the out-of-plane current distribution is qualitatively as discussed in Section \ref{review} and shown in Figure \ref{fig:qualitative3D}. To investigate the scaling, we plot the extent (full width at tenth maximum -- FWTM -- the maximum always being located at the null) of the $j_{y}$ distribution  along the $y$-axis (the component contributing to 3D reconnection along that fieldline) in Figure \ref{fig:dfit}. This measurement does not appear to follow a power-law scaling, perhaps due to the number of competing effects. We can understand the decrease in the FWTM as $\eta$ decreases as follows. At large $|y|$ there is a large magnetic pressure, that acts to halt the collapse. As $\eta$ is reduced, the current sheet -- in the absence of this magnetic pressure -- should become progressively thinner across its width-wise axis (see Figure \ref{fig:wlpeak}). However, for a given perturbation energy, the minimum current sheet width allowed by the magnetic pressure is fixed (though increases proportionally to $|y|$).
Therefore as $\eta$ is reduced, the point at which the magnetic back-pressure halts the collapse moves towards progressively smaller $|y|$, such that the collapse is only resistively halted in a small region around the null (in $y$), being halted by magnetic back-pressure for larger $|y|$. 
	 This is seen in the line plots to the right in Figure \ref{fig:dfit}.
	 All of the preceding implies that the current becomes increasingly peaked around $y=0$ as $\eta$ is decreased, leading to the observed scaling of the FWTM. 
\subsection{3D Reconnection rate, ${\eta}j$ at the null, and net flux transfer}

We next consider the peak reconnection rate attained during the collapse. For the 2D runs, the reconnection rate is defined simply by $\eta j_{y}$ at the null point. In three dimensions, the reconnection rate is the maximal value of
\begin{equation}\label{eq:recrate}
\int {\bf E}\cdot {\bf dl} = \int E_\|\, dl,
\end{equation}
the integration being performed along magnetic fieldlines {\citep{schindler1988}}. {Following \citet{2005GApFD..99...77P},} by symmetry this coincides in our simulations with $\int_{x=z=0}E_y\, dy$ (verified by calculating the integral (\ref{eq:recrate}) along $\sim \mathcal{O}(10^4)$ fieldlines traced from seed points clustered near the null).
	To compare the 2D and 3D rates, we simply multiply the recorded 2D rate by a factor  of $L_{y}=2$ (the domain size in $y$ for the 3D runs).
	These quantities are shown in Figure \ref{fig:rates}.
	For the 2D runs, we find in the linear case that the reconnection rate becomes independent of $\eta$ in this range, which is an expected scaling from \citet{1991ApJ...371L..41C,1993ApJ...405..207C} and was also found in the numerical experiments of \citet{1996ApJ...466..487M}. 
	However, contrary to the simulations of \citet{1996ApJ...466..487M} our 2D nonlinear peak reconnection does seem to exhibit some (albeit, weak) scaling, decreasing with $\eta$. 
	We hypothesise that this is due to the presence of Ohmic heating in our simulations (absent on those of \citealt{1996ApJ...466..487M}), which leads to an $\eta$-dependent increase in temperature, and therefore pressure, in the collapsing current layer. 

In the 3D simulations, we find that for both linear and nonlinear collapse the reconnection rate depends on both $\eta$ and the degree of magnetic field asymmetry, $k$. We note that the reconnection rate has a much stronger scaling with $\eta$ that the value of $\eta j$ at the null (compare the left and right frames of Figure \ref{fig:rates}). Thus, the strong scaling of the reconnection rate in 3D with $\eta$ is due primarily to the increasingly peaked current distribution along $y$ discussed in the previous section. That is, the depression of the 3D rate is mainly due to the strong scaling of $j_y$ with $\eta$ at large $y$, since it is essentially independent of $\eta$, $E_\|=\eta j_\|\sim \eta^{1}$. This directly determines affects the 3D reconnection rate, it being obtained by integrating $E_\|$ along $y$. As such, we find that the rotationally symmetric 3D null produces a smaller peak reconnection rates than the asymmetric nulls (with smaller $k$).

\section{Accessing fast reconnection through collapse in astrophysical plasmas}\label{sec:higherbeta}

\begin{figure*}
    \centering
    \subfigure[]{\includegraphics[width=0.4\linewidth]{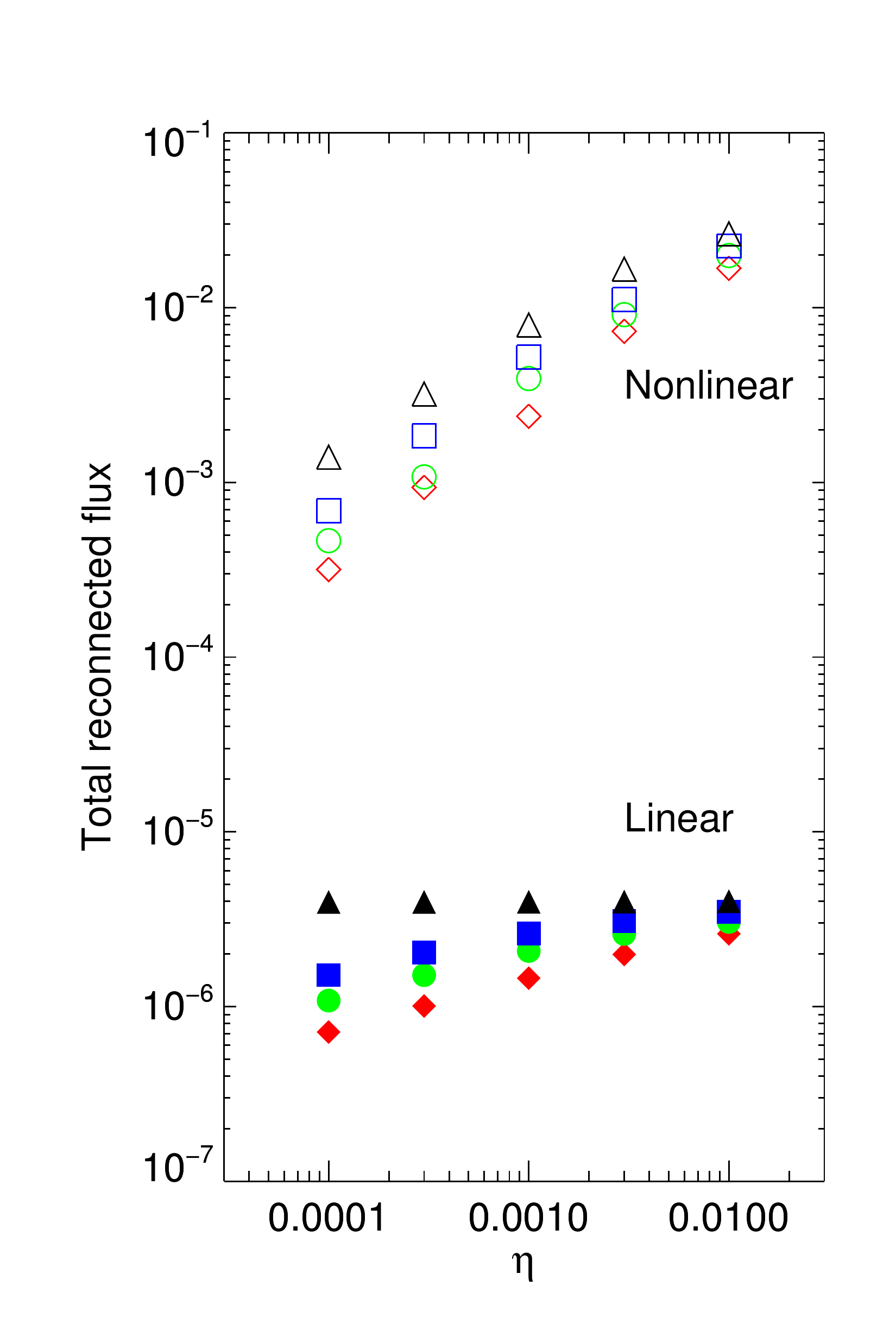} }
        \subfigure[]{\includegraphics[width=0.4\linewidth]{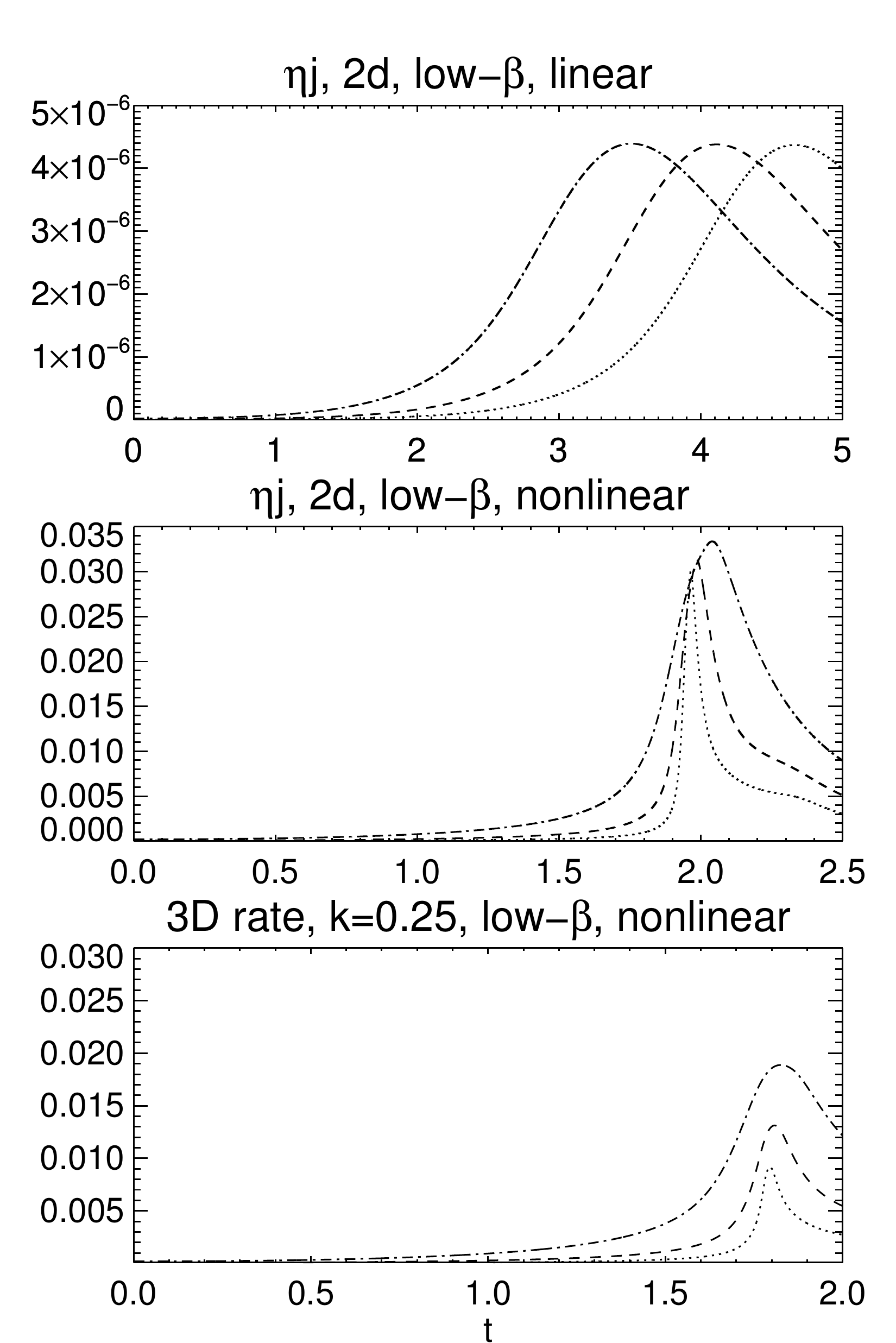} }

       \caption{Left: The total flux reconnected by the peak time in our low-$\beta$ parameter study Right: representative curves of the 2D and 3D reconnection rates in time achieved by the collapse for resistivity ${\eta}=10^{-4}$, $3\cdot10^{-4}$, and $10^{-3}$ for dashed-dot, dashed and dotted lines, respectively. .
         }      
 \label{fig:reconnectedflux}
\end{figure*}

\begin{figure*}
 \centering
    \includegraphics[width=0.875\linewidth]{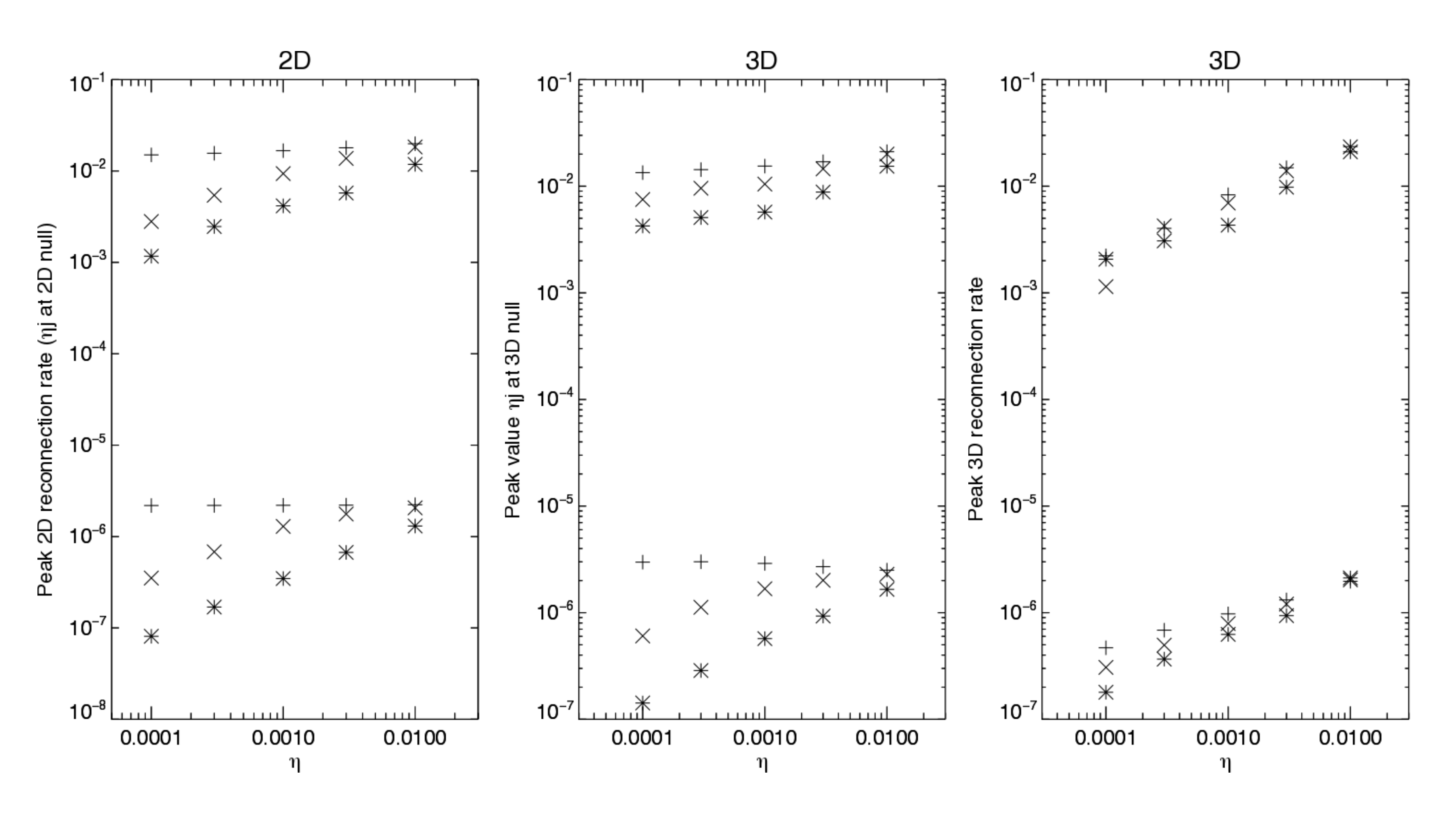} 
       \caption{Reconnection rate in the 2D simulations (left), ${\eta}j$ at the null in the $k=1$ 3D 
       simulations (center) and the corresponding true 3D reconnection rate (right) for variable ${\beta}=10^{-8},10^{-2},10^{-1}$, represented by $+$, $\times$ and ${\ast}$ symbols, respectively.  The linear ($j_{0}=10^{-5}$) and nonlinear  ($j_{0}=10^{-1}$) runs are simply distinguished as being the lower and upper cluster of points along the vertical axes, respectively, as the plotted quantities contain no normalisation by $j_{0}$.
       We find that in both cases a raised plasma-$\beta$ curtails current build up a the null (as the current sheet does not narrow as much), decreasing the reconnection rate in the 2D case. However, we find that in the 3D case, the null current makes only a small contribution to the overall reconnection rate. For $k=1$, the magnetic backpressure exerted by the $B_{y}$ component is sufficiently dominant mechanism curtailing collapse out of the $y=0$ plane for the $\beta$ considered. 
          }      
 \label{fig:highbeta}
\end{figure*}

\begin{figure}
 \centering
    \includegraphics[width=0.45\linewidth]{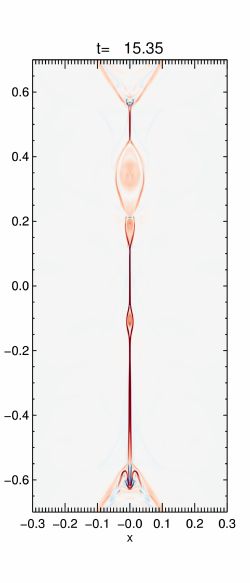} 
       \caption{Demonstration of nonlinear tearing in an unstable, high-aspect ratio current sheet produced by nonlinear null collapse. This particular setup is as per the 2D parameter study, but with larger amplitude and more appreciable ambient pressure ($\beta_{0}=10^{-2}$, ${\eta}=10^{-5}$ and $j_{0}=2$). The instability develops after the halting time of the initial collapse.
          }      
 \label{fig:plasmoid}
\end{figure}

	The main historical motivation of previous 2D collapse studies was assessing the viability of collapse as a fast reconnection mechanism, with a view towards, say, modelling flare energy release. 
	We have found in the previous section that the basic physics of the implosion for maintaining a favourable { (fast)} scaling of the product ${\eta}j$ is unchanged in 3D; however, the introduction of a spatially-dependent magnetic back-pressure in 3D ultimately limits the spatial extent of the non-ideal, high-${\eta}j$ region, which adversely impacts the reconnection rate. 
	It is currently unclear if this is a significant limiting factor of reconnection efficiency in real astrophysical applications involving 3D nulls.
	  This is in part due to the fact that the linear null employed here is a simplified model - in any real application the magnetic pressure would not grow monotonically away from the null. 
	  Therefore the impact of this stalling of the collapse will depend on the relative scales involved in the given problem.
	Nevertheless, we can conclude from our results that more (rotationally) asymmetric nulls (that are `closer to 2D') would generally be more favourable sites for magnetic energy release due to collapse than rotationally symmetric nulls. 

Regardless of the fast instantaneous reconnection rates reported, it has been pointed out by \citet{2000mare.book.....P} (Chapter 7) it is also important to consider the total  reconnected flux during the collapse in assessing the efficacy of magnetic energy release. Figure \ref{fig:reconnectedflux} shows the total flux reconnected up to the critical time for the various runs in our low-$\beta$ parameter study. We find that only in the linear 2D case is the amount of reconnected flux resistivity-independent. In this case, curves of ${\eta}j(t)$ exhibit a simple translation,  with the area underneath the curve remaining fixed. This translation is determined by the time required to advect the pulse at the background Alfv\'en speed to the diffusion-dominated scale (which is longer for smaller $\eta$). In the linear 3D case, this $\eta$-independence is no longer observed due to the action of the non-zero out of plane field component limiting the collapse at large $|y|$. The reconnected flux thus inherits the (relatively-weak) scaling with resistivity and the propensity to be lower as $k\rightarrow1$. For both the 2D and 3D {\emph nonlinear} collapses, however, the total flux reconnected by peak time is strongly dependent on resistivity due to the fact that the implosion only produces significantly enhanced ${\eta}j$  (the peak value of which is $\eta$-dependent) for a short duration in a highly-impulsive fashion, which can also be seen in the curves in Figure \ref{fig:reconnectedflux}. 

	An important further possible limitation of null collapse from the perspective of fast reconnection in astrophysical plasmas is the adverse effect of finite plasma pressure, first reported by \citet{1996ApJ...466..487M,2000mare.book.....P}. 
	The observation is that pressure increases in the imploding current sheet until a sufficiently large outward pressure grows, halting the collapse.  
	Given the nature of our results, there is no reason to expect this problem not to persist in the 3D case. 
	To verify this, we repeated the 2D simulations and the $k=1$ 3D simulations in the linear and nonlinear perturbation regimes for $\beta_{0}=10^{-2}$ and $\beta_{0}=10^{-1}$. 
	Figure \ref{fig:highbeta} shows the (2D) reconnection rate for the 2D simulations, the 3D reconnection rate for the 3D simulations, and the equivalent value of ${\eta}j$  at the null in the 3D simulations. 
	We see that, for the 2D case, the overall dependence of peak reconnection rate departs from (near)-$\eta$ independence in the low-$\beta$ case (horizontal line) and begins to scale more strongly, although the curves are always less steep than  a simple $\eta^{1}$ dependence which may be expected to be the limiting case for { a failed collapse where reconnection occurs at the static rate}.
 In the 3D, $k=1$ case, the current at the null itself behaves similarly however we note that there is not much impact on the overall 3D reconnection rate (as the current at the null itself only makes a small contribution to the reconnection rate). This is because the collapse at large $|y|$, as discussed in Section \ref{scaling},  is strongly limited by a magnetic back-pressure. This indicates that, for these nulls, this magnetic back-pressure, which here is 3D null equivalent of the  \lq{guide-field}\rq{} back-pressure discussed in the 2D literature, remains the dominant throttling mechanism. 
	Based upon the arguments given in \citet[][Chapter 7.1]{2000mare.book.....P}, for (linear 2D) collapse to not enter this { ambient} pressure-limited regime requires $\beta_{0}<{\eta}^{0.56}$, and so for solar resistivity of the order of ${\eta}\sim10^{-10}$  this requires $\beta_{0}{\le}10^{-6}$ on the boundary. Although there are inherent difficulties in ascribing real length scales to the boundaries of problems involving linear nulls, this value is smaller than typical $\beta$ inferred throughout various regions of the solar atmosphere via observation (however, other plasmas can have higher resistivities and therefore less prohibitive $\beta_{0}$ restriction).  
	Thus,  there remain doubts as to whether simple MHD null collapse is an efficient mechanism for magnetic energy release in the solar atmosphere, and we have found that considering the MHD collapse of a 3D null as opposed to 2D does not by itself remedy this problem. 

	The unfavourable scalings with resistivity in the nonlinear regime reported above thus suggest that the initial implosion may not provide a viable mechanism for fast energy release for coronal parameters. 		
	However, this initial collapse does set up a current sheet geometry that could lead to rapid energy release following some secondary process.
	Three candidates that we discuss below are secondary current sheet thinning as an MHD process, nonlinear tearing, or a collapse to collisionless scales. 
	First, a `secondary thinning' has been proposed by \citet{1996ApJ...466..487M}.
	The essential idea is that once the current sheet becomes highly pressurised, the pressure gradient may drive strong outflows that relieve the pressure enhancement, permitting further collapse. 	
	Accounting for thermal conductivity (excluded from our simulations) may also allow for a reduction in current sheet temperatures, also relieving internal current sheet pressure which sustains it against the inwardly directed Lorentz force of the implosion. 
	To our knowledge, secondary thinning processes have not yet been further considered, and we note that although it is tempting to suppose that this may occur more readily in 3D (due to the additional dimension for outflow), we actually have previously found that for these 3D nulls the plasma is predominantly ejected in a collimated jet near the plane of collapse due to the curvature of fieldlines during spine-fan reconnection \citep[][Figure 5]{2017ApJ...844....2T}.
	It is also the case that, if future attempts are made to investigate secondary thinning simulation, care should be taken that appropriate boundary conditions are employed (i.e.~those allowing the passage of outgoing waves), otherwise secondary thinning could in fact be caused by the reflection and return of the outgoing fast shocks rather than being a self-consistent process within the locality of the current sheet.  
	Alternatively, the collapse only needs to reach a scale which at which the effective local resistivity grows anomalously (i.e. a current-dependent resistivity), or a scale at which reconnection becomes collisionless \citep[e.g.][]{2007PhPl...14k2905T,2008PhPl...15j2902T}. 
	Otherwise, firmly within the realm of single-fluid, uniform resistivity MHD, it may be possible to promote fast reconnection via collapse { regardless of $\beta_{0}$} by having a sufficiently energetic collapse that  a current sheet forms with an aspect ratio that is susceptible to nonlinear tearing. This is now a well established route to fast reconnection \citep[e.g.][and references therein]{huang2017}, and has been demonstrated by occur in a modified form in 3D by \citet{daughton2011,wyper2014a}. At 3D nulls, it is expected that a nonlinear tearing of the current sheet occurs for Lundquist numbers greater than $2\times 10^{4}$ and current sheet aspect ratios greater than 100 \citep{wyper2014a}. Since the instability takes some time to set in, and the current sheet formed during our collapse simulations gradually broadens following the initial implosion, it may be that the current layer at the critical time should have a larger aspect than this for nonlinear tearing to set in; however, this requires more careful study in a simulation with `open' boundaries. What is clear is that -- examining the nonlinear scalings for $w$ and $l$ obtained from Figure \ref{fig:wlpeak}, for a given value of $\eta$, a sufficiently energetic perturbation should yield a current sheet beyond the critical aspect ratio.
We have ourselves been able to observe current sheet tearing in 2D collapses after the initial phase of collapse (Figure \ref{fig:plasmoid}). In this simulation, we consider the same 2D setup as in the parameter study but with $\beta_{0}=10^{-2}$, $\eta=10^{-5}$, and $j_{0}=2$  (larger than perturbations previously considered). The same nonlinear implosion process proceeds, rapidly producing a high-aspect ratio current sheet by $t_{peak}\approx{0.6}$ which subsequently undergoes nonlinear tearing.
	This particular simulation however suffers from questions relating to the applicability of the closed boundary, namely that the current sheet has been impacted upon by the reflected fast waves before the instability develops have and also that, once it does, ejected plasmoids are artificially confined near the boundary due to the no-flow through boundary conditions. 	
	At present, it is unclear whether these effects of non-physical confinement significantly effect the evolution of the instability (and most crucially, whether it occurs at all). 
	As such, we caution that this is a preliminary result intended primarily as a conceptual demonstration, and hope to study this further in the future (ideally, with an open system).

\section{Conclusions and Discussion}\label{discussion}

	In this paper we have considered a detailed parameter study of collapsing 3D magnetic null points of variable eccentricity ($k=0.25,0.5,1$), alongside 2D nulls (equivalent to $k=0$) with variable resistivity and variable initial perturbation amplitude, for both low and moderate plasma-$\beta$.   The Key Findings are as follows:

\begin{enumerate}
\item The collapse of 3D nulls, across variable-$k$, is found to be both \textit{qualitatively} and \textit{quantitatively} as per the 2D case within the \textit{plane of collapse} (here, the $y=0$ plane). In both 2D and 3D, there exist two regimes of collapse: linear and nonlinear. These are characterised by self-similar and quasi-1D evolution, respectively, where the nature of the implosion is dependent on the relative energy of the perturbation and ability of the plasma to diffuse or resist the perturbation. For both regimes we find that the implosion proceeds to increasingly small length-scales as resistivity is decreased (here, seen and measured as the achieved current sheet widths, Figure \ref{fig:wlpeak}). This length scale, which determines the current density via $\mathbf{\nabla}\times\mathbf{B}$, is found to scale with resistivity as ${\eta}^{0.5}$ (linear) and ${\eta}^{0.89}$ (nonlinear). 
This leads to an independent (linear) or  weak dependence (nonlinear) of the product ${\eta}j$ at the null on the resistivity, which is the reconnection rate in 2D rate and also contributes to the 3D rate. 

\item The crucial difference between 2D and 3D null collapse occurs out of the $y=0$ plane. For $k>0$, the magnetic field component ($B_{y}=ky$) increases away from the $xz$-plane in which the collapse proceeds. 
	This field component acts to provide a magnetic back-pressure that grows with distance from the null, opposing the collapse, and is analogous to the effect of guide-field back-pressure considered in 2D by \citet{1996ApJ...466..487M}. 
	Thus, in the third dimension, the growth of ${\eta}j$ for $y\neq0$ is increasingly limited for more rotationally symmetric 3D null points ($k\rightarrow1$). 
	As the 3D reconnection rate (equation \ref{eq:recrate}) is determined by the parallel electric field throughout the non-ideal volume (as opposed to ${\eta}j$ at the null itself), this magnetic back-pressure limits the overall reconnection rate for 3D nulls and so we find that the overall reconnection rate  appreciably decreases as $\eta$ is reduced, despite the favourable scalings for ${\eta}j$ at the null itself discussed in Key Finding 1  (cf. the left and right panels of Figure \ref{fig:rates}). 

\item 
Increasing the plasma-$\beta$ to realistic values for the solar corona ($\beta_0\sim 10^{-2}-10^{-1}$) inhibits the collapse, leading to reduced reconnection rates. This effect is less severe in 3D since the magnetic back-pressure tends to be the dominant throttling mechanism across the majority of the current layer volume. 

\end{enumerate}

There are a number of astrophysical implications of these findings. Firstly, from the perspective of low-$\beta$ null collapse as a fast reconnection mechanism, 
although we find that the basic physics of the implosion for maintaining favourable scalings of the product ${\eta}j$ is unchanged in 3D (Key Finding 1), it appears that a further complication of 3D collapse is that the introduction of a spatially-dependent magnetic back-pressure (Key Finding 2) ultimately limits the spatial extent of the non-ideal, high-${\eta}j$ region, which negatively impacts upon the reconnection. Thus, we can conclude from our results that more eccentric, rotationally asymmetric nulls that would generally be more favourable in terms of magnetic energy release than a more rotationally symmetric counterpart. 
Regardless, 
we show directly that even if the peak reconnection rate obtains a favourable scaling with resistivity, that the total reconnected flux by the critical time is still limited by decreasing resistivity (Figure \ref{fig:reconnectedflux}). It may be the case that higher rates of reconnection are maintained after the stalling of the collapse, but we cannot directly consider this from our simulations for the parameter study. 
As $\beta_{0}$  (the value of $\beta$ at the boundary) is raised to values that are thought to be representative of the solar atmosphere, we have found that in the case of 3D null collapse the increasing initial plasma pressure interferes with the favourable scaling of the reconnection rate as resistivity is lowered (Key Finding 3, Section \ref{sec:higherbeta}). 
	Thus,  there remain old and new doubts as to whether simple MHD null collapse is an efficient mechanism for magnetic energy release in the solar atmosphere, and we have found that considering the MHD collapse of a 3D null as opposed to 2D does not by itself provide any remedy to this problem.
	However, in Section \ref{sec:higherbeta} we have discussed several ways in which null collapse could bring about efficient reconnection and magnetic energy release as a secondary process even if the scaling of the initial reconnection rate itself is limited (either by the 3D out-of-plane magnetic pressure, or by too-high ambient plasma pressures).
	These processes include secondary thinning, accessing sufficient scales for current-dependent resistivity, microphysics and collisionless reconnection, and finally, by creating a current sheet that is susceptible to tearing after the implosion. We have been able to demonstrate tearing as the result of a 2D collapse using the simulation setup considered in these papers for relatively high $\beta_{0}=10^{-2}$ for the case of a closed system.

	Outside of the fast reconnection perspective, we were also motivated to investigate 3D null collapse as we expect the results to be useful in future theoretical studies of \textit{Oscillatory Reconnection} (OR).
	 OR  \citep{2009A&A...493..227M,2017ApJ...844....2T} is a time-dependent and oscillatory magnetic reconnection system which is currently considered as a candidate for explaining \textit{quasi-periodic pulsations} (QPPs) in solar and stellar flares \citep{2009SSRv..149..119N,2016SoPh..tmp..147V,McLaughlin2018}, where a crucial question in assessing its applicability is;  \textit{what periods can it produce for solar flare parameters (and are they compatible with QPPs)?} It has been shown that the period of OR is  dependent upon the initial disturbance to the null point field, behaving akin to a damped harmonic oscillator \citep{2012A&A...548A..98M}, and therefore a deeper understanding of null collapse can enhance our understanding of OR. This may be achieved by either simple assumptions regarding the proportion of available wave energy reaching close to the null, which would be equated to the null collapse system considered here as the initial perturbation energy, or with complementary  numerical modelling efforts to estimate how much wave energy makes it to the immediate vicinity of the null whilst accounting for effects such as mode-conversion and atmospheric stratification \citep[e.g.,][]{2017ApJ...837...94T}. 
Indeed, our results could be used in tandem with global-scale simulations of these processes, which by necessity cannot resolve the details of the current sheet and reconnection dynamics.

	Finally, we note that it may be possible to investigate the processes of implosive current sheet formation as described in this paper (and its predecessors) in the laboratory with devices such as CS-3D  \citep{0741-3335-41-3A-062,Frank2001,2017PlPhR..43..696F}, which can investigate current sheet formation and subsequent reconnection in a variety of null-containing fields including null lines (2D null points), null lines with guide fields, and truly 3D nulls. 
	In these experiments a over-dense current sheet forms after some time delay which correlates with times expected for radial  propagation of converging magnetoacousic waves from the edge to the centre of the chamber, which is then proceeded by {\lq metastable}\rq{} stage, which is then followed by an eventual \lq{}explosive\rq{} release of magnetic energy. 
	We suggest that the theory of null collapse may describe the physics of the initial current sheet formation within such devices, explaining the compression ratios and sheet thickness achieved. 
	Indeed, some of the reported results are suggestive of behaviour as predicted by null collapse theory, for example, it has been observed that electrical current is less effectively concentrated at small scales (i.e. measured current sheets are thicker) and the plasma within is less effectively compressed in response to growing guide fields in the case of the 2D null line with guide field \citep{2005PhPl...12e2316F}.

\section*{Acknowledgements}

The authors acknowledges generous support from the  Leverhulme Trust and this work was funded by a Leverhulme Trust Research Project Grant: RPG-2015-075. The authors acknowledge IDL support provided by STFC. The computational work for this paper was carried out on HPC facilities provided by the Faculty of Engineering and Environment, Northumbria University, UK.  JAM acknowledges STFC for support via ST/L006243/1. DIP acknowledges STFC for support via ST/N000714/1.

\bibliography{references}

\appendix
\section{Nondimensionalisation and the Solver (LareXd code)}\label{appendixA}
{
Following the details  in the LareXd user manual,  the normalisation is through the choice of three basic normalising constants, specifically:
\begin{eqnarray*}
x&=&L_0 \hat{x}\\
\mathbf{B}&=&B_0\hat{\mathbf{B}} \\
\rho&=&\rho_0 \hat{\rho}
\end{eqnarray*}
where quantities with and without a hat symbol are dimensional and nondimensional, respectively. These  are then used to define the normalisation of quantities with derived units through
\begin{eqnarray*}
v_{0}&=&\frac{B_{0}}{\sqrt{\mu_{0}\rho_{0}}}\\
P_{0}&=&\frac{B^{2}_{0}}{\mu_{0}} \\
t_0&=&\frac{L_0}{v_0}\\
j_{0}&=&\frac{B_{0}}{\mu_{0}L_{0}}\\
E_0&=&v_0 B_0\\
\epsilon_0&=&v_0^2
\end{eqnarray*}
so that $\mathbf{v}=v_0\hat{\mathbf{v}}$, $\mathbf{j}=j_0\hat{\mathbf{j}}$, $t=t_0\hat{t}$ and 
$P=P_0\hat{P}$ etc. 
Applying this normalisation to the ideal MHD equations simply removes the vacuum 
permeability $\mu_0$. In resistive MHD, this scheme leads naturally to a resistivity normalisation:
 \begin{displaymath}
 \hat{\eta}=\frac{\eta}{\mu_0 L_0 v_0}
\end{displaymath}
or $\eta_0=\mu_0 L_0 v_0$. Since $v_0$ is the normalised Alfv\'en speed this means that 
$\hat{\eta}=1/S$ where $S$ is the Lundquist number as defined by the basic normalisation constants. 
}

The simulation is the numerical solution of the nondimensional, resistive MHD equations: ({NB: we drop the hat from this point onwards in the appendix, and throughout the main paper all quantities are nondimensional)}
\begin{eqnarray}
\frac{\mathrm{D}\rho}{\mathrm{D}t}&=&-\rho \nabla\cdot \mathbf{v}\\
\frac{\mathrm{D}\mathbf{v}}{Dt}&=&\frac{1}{\rho}(\nabla\times\mathbf{B})\times\mathbf{B}
-\frac{1}{\rho}\nabla p + \mathbf{F}_{shock}\\
\frac{\mathrm{D}\mathbf{B}}{\mathrm{D}t}&=&(\mathbf{B}\cdot\nabla)\mathbf{v}-\mathbf{B}
(\nabla\cdot\mathbf{v})-\nabla\times(\eta\nabla\times\mathbf{B})\\
\frac{\mathrm{D}\epsilon}{\mathrm{D}t}&=&-\frac{p}{\rho}\nabla\cdot\mathbf{v}+\frac
{\eta}{\rho}j^{2} + \frac{\mathbf{H}_{visc}}{\rho}\\
\mathbf{j} &=& \mathbf{\nabla}\times\mathbf{B}\\
\mathbf{E} &=& -\mathbf{v}\times\mathbf{B}+\eta\mathbf{j}\\
p &=& \epsilon\rho\left(\gamma-1\right)
\end{eqnarray}
which are solved on a Cartesian grid using the \textit{Lare2d} and \textit{Lare3d} codes. All results presented are in non-dimensional units. Algorithmically, the code solves the ideal MHD equations explicitly using a Lagrangian remap approach and includes the resistive terms using explicit subcycling \citep{2001JCoPh.171..151A,2016ApJ...817...94A}. The solution is fully nonlinear and captures shocks via an edge-centred artificial viscosity approach \citep{1998JCoPh.144...70C}, where shock viscosity is applied to the momentum equation through $\mathbf{F}_{shock}$ and heats the system through $\mathbf{H}_{visc}$. Extended MHD options available within the code, such as the inclusion of Hall terms, were not used in these simulations. Full details of the code can be found in the original paper \citep{2001JCoPh.171..151A} and the users manual.

\section{Boundary conditions}\label{boundary}

	The calculations presented in this paper represent the solution for the case of perturbed nulls contained within the cube $|x,y,z|\le1$. 
	The faces are subject to boundary conditions that permit no flow through or along the boundary ($\mathbf{v}=\mathbf{0}$) with zero-gradient conditions taken $\rho$ and $\epsilon$, and also on magnetic field components which are tangential to a given face. The normal component of the field is held fixed (line-tied) through the boundary. The suitability of these boundary conditions, and overall stability of the setup, was checked by runs with no perturbation where we found that there was no undesirable behaviour such as the launching of spurious waves or erroneous current formation at the boundary. 
	In practice, we apply the aforementioned conditions only on `external' computational boundaries, and exploit appropriate symmetry/antisymmetry conditions on the `internal'  computational boundaries. 
	Specifically, for the 3D setup we solve  only for  the half-cube $y\ge0$, and in 2D for the quarter-plane   $0\le(x,z)\le1$ ($y=0$, arbitrarily).
	The more favourable reduction in 2D is facilitated by the fact that the form of equation (\ref{eq:2dnull}) results in the current sheets length-wise and width-wise axes aligning with a computational boundary, hence why we have rotated the field in the 2D case. 
	The implementation and accuracy of the symmetry conditions were checked simply by re-running some simulations in the whole domain, and we find perfect agreement.
	We note that the symmetry conditions are not used for the tearing mode simulation (Figure \ref{fig:plasmoid}), but rather we simulate the full domain order to permit the symmetry breaking expected to occur during the instability.

\section{Grid geometry, resolution and testing}

	To adequately resolve the small-scale features produced by the collapse, especially in the lower resistivity cases, grid stretching is employed to concentrate resolution in the vicinity of the current sheets. 
	The grid is stretched according to a variation on the scheme of \citet{1971LNP.....8..171R}, namely the cell boundary positions $x_{b}$ along the $x$-direction  are distributed according the transformation:
\begin{eqnarray}
x_{b} &=&  \left\lbrace 1 + \frac{\sinh\left[\lambda\left(\xi_{i}-\Gamma\right)\right]}{\sinh\left[\lambda\Gamma\right]} \right\rbrace - 1 \\
\Gamma &=& \frac{1}{2\lambda}\ln\left[\frac{1+\left(e^{\lambda}-1\right)0.5}{1-\left(1-e^{-\lambda}\right)0.5}\right]
\end{eqnarray}
where $\xi_{i}$ is a uniformly distributed computational coordinate $\xi\in[0,1]$ subdivided amongst the number of cells used in the $x$ direction. The degree of grid clustering at the origin (the null point) is controlled by the stretching parameter $\lambda$. Likewise, the same form and parameters are used for the distribution of cells in $y$ and $z$. 
	In our final 3D simulations of the parameter study, presented here we chose $\lambda = 8$ then performed simulations with increasing numbers of cells up to a maximum of $nx=nz=1024$, $ny=512$ (effectively, $1024^3$ given the symmetry). 
	Generally, we found that provided the  resolution is sufficient to stop the current sheet collapsing to the grid-scale (i.e., capture the physics of the pressurisation and resistive heating of the current sheet which facilitates the halting process) the solution as measured by the maximal values of current density, density and other variables at the null itself demonstrates convergent behaviour as the numerical resolution is increased. 
	In practice, only the simulations for the smallest resistivity can be ran within a reasonable time at $1024^3$, due to the unfavourable impact of smaller cell sizes upon the resistive timestep (${\Delta}t_{\eta}\propto\Delta_{x}/\eta$). Conveniently, however, higher values of $\eta$ correspond to much wider current sheets at the critical time which therefore do not require such a fine grid (see Figure \ref{fig:wlpeak}). 
	The final resolution as used for the data presented in the parameter study is as follows (all k, linear and nonlinear);  $10^{-4}\le{\eta}\le10^{-3}$ use $1024^3$ cells yielding ${\Delta}x_{min}\approx0.00029$,  ${\eta}=3\times10^{-3}$ uses $512^3$ cells  yielding ${\Delta}x_{min}\approx0.00058$, and $\eta=10^{-2}$ uses $256^3$ cells, yielding ${\Delta}x_{min}\approx0.00115$. Each of these final simulations is in good agreement with a simulation at half the stated resolution (half of the cells in each dimension), in a qualitative sense during the evolution of the implosion and in the sense of producing the same scaling laws (which are in agreement with analytical results). 
	They are also in an acceptable  level of quantitative agreement with lower resolution simulations, with difference in measured current at the null being less than one percent when compared to the half-resolution case, except in the case of  $\eta=10^{-4}$  (the most challenging to resolve) which has the largest difference ($\sim3\%$). Overall, given the excellent agreement with analytic results for collapse scaling, which provide an independent means of verification where applicable, we are confident our simulations have faithfully captured the key aspects of the collapse up to the critical time. In 2D, equivalent stretching schemes are utilised in $x$ and $z$, however in test runs we also accessed much higher resolutions than possible for 3D for the sake of further testing (similar tests are also performed regarding variable stretching factors $\lambda$, to test the stretching).


\end{document}